\documentclass[reprint,tightenlines,showpacs,amsmath,amssymb]{revtex4-2}

\usepackage[T1]{fontenc}
\usepackage{graphicx}
\usepackage{dcolumn}
\usepackage{bm}
\usepackage{rotating}
\usepackage{siunitx}
\usepackage{placeins}
\usepackage{hyperref}
\hypersetup{
    colorlinks=true,
    linkcolor=blue,
    filecolor=magenta,      
    urlcolor=cyan,
    }

\begin{document}
\normalsize
\parskip = 5pt plus 1pt minus 1pt

\title{\boldmath Search for $X(3872)\to\pi^0\chi_{c0}$ and $X(3872)\to\pi\pi\chi_{c0}$ at BESIII}

\author{
\begin{small}
\begin{center}
M.~Ablikim$^{1}$, M.~N.~Achasov$^{10,b}$, P.~Adlarson$^{67}$, M.~Albrecht$^{4}$, R.~Aliberti$^{28}$, A.~Amoroso$^{66A,66C}$, M.~R.~An$^{32}$, Q.~An$^{63,50}$, X.~H.~Bai$^{58}$, Y.~Bai$^{49}$
% ...
Yi2020~Wang$^{53}$, Z.~Wang$^{1,50}$, Z.~Y.~Wang$^{1,55}$, Ziyi~Wang$^{55}$, D.~H.~Wei$^{12}$, F.~Weidner$^{60}$, S.~P.~Wen$^{1}$, D.~J.~White$^{59}$, U.~Wiedner$^{4}$, G.~Wilkinson$^{61}$
% ...
\\
\vspace{0.2cm}
(BESIII Collaboration)\\
\vspace{0.2cm} {\it
$^{1}$ Institute of High Energy Physics, Beijing 100049, People's Republic of China\\
% ...
$^{4}$ Bochum Ruhr-University, D-44780 Bochum, Germany\\
% ...
$^{10}$ G.I. Budker Institute of Nuclear Physics SB RAS (BINP), Novosibirsk 630090, Russia\\
% ...
$^{12}$ Guangxi Normal University, Guilin 541004, People's Republic of China\\
% ...
$^{28}$ Johannes Gutenberg University of Mainz, Johann-Joachim-Becher-Weg 45, D-55099 Mainz, Germany\\
% ...
$^{32}$ Liaoning Normal University, Dalian 116029, People's Republic of China\\
% ...
$^{49}$ Southeast University, Nanjing 211100, People's Republic of China\\
$^{50}$ State Key Laboratory of Particle Detection and Electronics, Beijing 100049, Hefei 230026, People's Republic of China\\
% ...
$^{53}$ Tsinghua University, Beijing 100084, People's Republic of China\\
% ...
$^{55}$ University of Chinese Academy of Sciences, Beijing 100049, People's Republic of China\\
% ...
$^{58}$ University of Jinan, Jinan 250022, People's Republic of China\\
$^{59}$ University of Manchester, Oxford Road, Manchester, M13 9PL, United Kingdom\\
$^{60}$ University of Muenster, Wilhelm-Klemm-Str. 9, 48149 Muenster, Germany\\
$^{61}$ University of Oxford, Keble Rd, Oxford, UK OX13RH\\
% ...
$^{63}$ University of Science and Technology of China, Hefei 230026, People's Republic of China\\
% ...
$^{66}$ University of Turin and INFN, (A)University of Turin, I-10125, Turin, Italy; (B)University of Eastern Piedmont, I-15121, Alessandria, Italy; (C)INFN, I-10125, Turin, Italy\\
$^{67}$ Uppsala University, Box 516, SE-75120 Uppsala, Sweden\\
% ...
$^{b}$ Also at the Novosibirsk State University, Novosibirsk, 630090, Russia\\
% ...
}
\end{center}
\vspace{0.4cm}
\end{small}
}

\date{April 4, 2022}

\newcommand{\RMAIN}{$\frac{\mathcal{B}(X(3872)\rightarrow\pi^0\chi_{c0})}{\mathcal{B}(X(3872)\to\pi^+\pi^-J/\psi)}$}
\newcommand{\RRYAN}{$\frac{\mathcal{B}(X(3872)\rightarrow\pi^0\chi_{c1})}{\mathcal{B}(X(3872)\to\pi^+\pi^-J/\psi)}$}
\newcommand{\RDOUBLE}{$\frac{\mathcal{B}(X(3872)\rightarrow\pi^0\chi_{c0})}{\mathcal{B}(X(3872)\to\pi^0\chi_{c1})}$}
\newcommand{\RTWOPI}{$\frac{\mathcal{B}(X(3872)\rightarrow\pi^+\pi^-\chi_{c0})}{\mathcal{B}(X(3872)\to\pi^+\pi^-J/\psi)}$}
\newcommand{\RTWOPIZERO}{$\frac{\mathcal{B}(X(3872)\rightarrow\pi^0\pi^0\chi_{c0})}{\mathcal{B}(X(3872)\to\pi^+\pi^-J/\psi)}$}

\newcommand{\NORMEFF}{32.5\%}
\newcommand{\NORMNUMBER}{$88.7^{+10.4}_{-9.7}$}

\newcommand{\SIGNIFMIN}{1.3$\sigma$}

\newcommand{\NORMEFFUNC}{32.5$\pm$0.2\%}

\newcommand{\BFUNCONE}{$4.7\%$}
\newcommand{\BFUNCTWO}{$10.8\%$}
\newcommand{\BFUNCTHREE}{$14.1\%$}

\newcommand{\SYSUNC}{$12.8\%$}
\newcommand{\FINALSYSUNC}{$14.5\%$}
\newcommand{\PIPISYSUNC}{$12.3\%$}
\newcommand{\TWOPIZEROSYSUNC}{$16.4\%$}

\newcommand{\SIGNIF}{$2.4\sigma$}
\newcommand{\UL}{3.6}

\newcommand{\FINALUL}{4.5}

\newcommand{\PIPIUL}{0.56}

\newcommand{\TWOPIZEROUL}{1.7}

\begin{abstract}
  Using 9.9 fb$^{-1}$ of $e^+e^-$ collision data collected by the BESIII detector at center-of-mass energies between 4.15 and 4.30 GeV, we search for the processes $e^+e^-\to\gamma X(3872)$ with $X(3872)\rightarrow\pi^0\chi_{c0}$ and $X(3872)\rightarrow\pi\pi\chi_{c0}$. We set upper limits (at 90\% C.L.) of \RMAIN$<\UL$, \RTWOPI$<\PIPIUL$, and \RTWOPIZERO$<\TWOPIZEROUL$. Combined with the BESIII measurement of $X(3872)\to\pi^0\chi_{c1}$, we also set an upper limit of \RDOUBLE$<\FINALUL$.
\end{abstract}

\maketitle


\section{\label{sec:level1} Introduction}

The $X(3872)$, discovered by the Belle experiment in 2003~\cite{firstObs}, was the first state in the charmonium region that could not be easily explained by a simple $c\bar{c}$ model. Despite having quantum numbers $J^{PC}=1^{++}$~\cite{numbers} and a mass near the predicted $\chi_{c1}(2P)$ mass of 3.95~GeV/$c^2$~\cite{chic}, the state has several properties that cannot be explained by a pure charmonium state above open charm threshold. The state is exceptionally narrow, with a measured width of $0.96^{+0.19}_{-0.18}\pm 0.21$~MeV~\cite{narrow1} or $1.39\pm 0.24\pm 0.10$~MeV~\cite{narrow2}, depending on the assumed lineshape. In addition, the state has large isospin violation effects in its decays, which is clearly seen in the fact that the decays $X(3872)\to\rho J/\psi$ and $X(3872)\to \omega J/\psi$ occur at approximately the same rate. Many $X(3872)$ decay modes have been observed, including $\rho J/\psi$~\cite{rho}, $D^0\bar{D}^{*0}$~\cite{dd}, $\gamma J/\psi$~\cite{gamma}, $\pi^0\chi_{c1}$~\cite{ryan} and $\omega J/\psi$~\cite{besX}. However the nature of the $X(3872)$ remains unclear. Since the mass of this state is near $D^0\bar{D}^{*0}$ threshold~\cite{pdg}, one explanation of its exotic properties is that the state has a $D^0\bar{D}^{*0}$ molecular component.

\begin{table*}[!ht]
  \caption{Theoretical predictions on \RMAIN$\textrm{ }$and \RDOUBLE$\textrm{ }$for different physical interpretations of the $X(3872)$ state. }
\begin{tabular}{ |c|c|c|c|c| }
\hline
 Ref & Technique & Interpretation &  \RMAIN & \RDOUBLE \\ \hline
~\cite{voloshin} & Multipole expansion & Four-quark/molecule & NA & 2.97 \\ \hline
~\cite{voloshin} & Multipole expansion& $\chi_{c1}(2P)$ & 0.0 & 0.0 \\ \hline
~\cite{mehen} & Effective field theory & $D^0\overline{D}^{0*}$ & NA & 2.84 to 2.98 \\ \hline
~\cite{wu} & Effective field theory & $D^0\overline{D}^{0*}+D^+D^{-*}$ & 1.3 to 2.07 & 1.65 to 1.77 \\\hline
~\cite{dong} & Effective field theory & $D^0\overline{D}^{0*}+D^+D^{-*}$ & NA & 3.72\\\hline
~\cite{zhou} & Effective field theory & $D^0\overline{D}^{0*}+D^+D^{-*}$+$\chi_{c1}(2P)$ & 0.094 & 1.15 \\ \hline
\end{tabular}

\label{tab:theory}
\end{table*}

By searching for new decay modes, we can learn about the quark configuration of the $X(3872)$ state. The ratios \RMAIN$\textrm{ }$and \RDOUBLE$\textrm{ }$ are expected to be sensitive to different physical interpretations of the $X(3872)$. Theoretical predictions under the hypothesis of a pure charmonium state~\cite{voloshin}, a generic four quark state~\cite{voloshin}, a $D^0\bar{D}^{0*}$ molecule~\cite{mehen}, a $D\bar{D}^*$ molecule with charged and neutral components~\cite{wu,dong}, or a combination of a molecular state and charmonium state~\cite{zhou} are  summarized in Table~\ref{tab:theory}. Since BESIII recently measured \RRYAN$ = 0.88^{+0.33}_{-0.27}\pm 0.10$~\cite{ryan}, we can measure both of these quantities at BESIII. The Belle Collaboration has also performed a search for $X(3872)\to\pi^0\chi_{c1}$, and they set an upper limit on \RRYAN$\textrm{ }$of 0.97 at 90\% confidence level~\cite{belleSearch}, which is consistent with the measurement of BESIII~\cite{ryan}.

In this paper, we search for the process $e^+e^-\to\gamma X(3872)$ with $X(3872)\to\pi^0\chi_{c0}$ and the $\chi_{c0}$ decaying hadronically to the final states shown in Table~\ref{tab:chicDecays}. These final states are chosen because they have large branching fractions and can be reconstructed with a high efficiency. In addition, we search for the double pion transitions $X(3872)\to\pi^+\pi^-\chi_{c0}$ and $X(3872)\to\pi^0\pi^0\chi_{c0}$ through the same hadronic decays of the $\chi_{c0}$. Based on a molecular interpretation for the $X(3872)$, effective field theory (EFT) calculations predict that $\frac{\mathcal{B}(X(3872)\to\pi\pi\chi_{c0})}{\mathcal{B}(X(3872)\to\pi^0\chi_{c0})}\approx\mathcal{O}(10^{-3})$~\cite{dong} or $\mathcal{O}(10^{-5})$~\cite{mehen}, depending on the specific methods used.

\begin{table}[h]
  \caption{ Hadronic decay modes of the $\chi_{c0}$ reconstructed in this analysis, with branching fractions from Ref. \cite{pdg}.}
  \centering
\begin{tabular}{ |c|c| }
\hline
 Decay & Branching Fraction (\%) \\ \hline
 $\chi_{c0}\to\pi^+ \pi^-$ & $0.567\pm0.022$ \\ \hline
 $\chi_{c0}\to K^+ K^-$ & $0.605\pm0.031$ \\ \hline
 $\chi_{c0}\to\pi^+\pi^-\pi^+\pi^-$ & $2.34\pm 0.18$ \\ \hline
 $\chi_{c0}\to\pi^+\pi^- K^+ K^-$ & $1.81\pm0.14$\\ \hline
 $\chi_{c0}\to\pi^+\pi^-\pi^0 \pi^0$ & $3.3\pm0.4$\\ \hline
\end{tabular}
\label{tab:chicDecays}
\end{table}

We normalize our results to the process $e^+e^-\to\gamma X(3872)$ with $X(3872)\to\pi^+\pi^-J/\psi$ with $J/\psi\to\ell^+\ell^-$, for $\ell=e$ or $\ell=\mu$. This normalization channel is chosen because it has large statistics and is easy to reconstruct. We measure the ratio \RDOUBLE$\textrm{ }$by using the efficiency and fit results for $X(3872)\to\pi^0\chi_{c1}$ from Ref.~\cite{ryan}. In all cases, the $X(3872)$ is produced through $e^+e^-\to\gamma X(3872)$. Since this is always the same for the search and normalization channels, the production cross section cancels in the ratio, so we can combine data sets from different energy points in our analysis. The normalizations also cancel several systematic uncertainties.

\section{Experimental Background}

The BESIII experiment, operating at the Beijing Electron Positron Collider (BEPCII), has measured $\sigma(e^+e^-\to\gamma X(3872))\mathcal{B}(X(3872)\to\pi^+\pi^-J/\psi)$ for center-of-mass energies ($E_{\rm CM}$) between 4.008 and 4.60 GeV and found that the cross section is the largest for 4.15$<E_{\rm CM}<$4.3 GeV~\cite{besX}. BESIII has 9.9~fb$^{-1}$ of data in this energy region, at the energies shown in Table~\ref{tab:signalData}. Center-of-mass energies are measured in Refs.~\cite{ecm1} and~\cite{ecm2}, while the luminosities are measured in Refs.~\cite{lum1} and~\cite{lumscan}. This data set makes it possible to search for complicated decay modes of the $X(3872)$ state.

\begin{table}[ht]
  \caption{Data with $4.15<E_{\rm CM}<4.3$~GeV, where the $X(3872)$ is produced via $e^+e^-\to\gamma X(3872)$. All of these energy points are used in our nominal fit. The total luminosity is 9.9~fb$^{-1}$. Only statistical uncertainties are shown. Entries that do not include a reference are estimated values. The last line of the table shows the range of luminosity values for 29 different energy points that each have a much smaller luminosity than the other data points.}
  \centering
  \begin{tabular}{ |r@{}l|c|c| }
\hline
 \multicolumn{2}{|c|@{}}{Luminosity (pb$^{-1}$)} & $E_{\rm CM}$ (MeV) & Year  \\ \hline
 \multicolumn{2}{|c|}{401.5} & 4157.83$\pm0.05$~\cite{ecm2} & 2019  \\\hline
 \multicolumn{2}{|c|}{3189.0}  & 4178 & 2016  \\ \hline
 43.09&$\pm$0.03~\cite{lum1} & 4188.59$\pm 0.15$~\cite{ecm1} & 2013  \\ \hline
 526.70&$\pm 2.16$~\cite{ecm2}  & 4189.12$\pm0.05$~\cite{ecm2} & 2017 \\ \hline
 526.60&$\pm2.05$~\cite{ecm2}  & 4199.15$\pm0.06$~\cite{ecm2} & 2017 \\ \hline
 54.55&$\pm$0.03~\cite{lum1} & 4207.73$\pm 0.14$~\cite{ecm1} & 2013  \\ \hline
 517.10&$\pm1.81$~\cite{ecm2}  & 4209.39$\pm0.06$~\cite{ecm2}& 2017 \\ \hline
 54.13&$\pm$0.03~\cite{lum1} & 4217.13$\pm0.14$~\cite{ecm1} & 2013  \\ \hline
  514.60&$\pm1.80$~\cite{ecm2}  & 4218.83$\pm0.06$~\cite{ecm2} & 2017 \\ \hline
 1047.34&$\pm$0.14~\cite{lum1} & 4226.26$\pm0.04$~\cite{ecm1} & 2013 \\ \hline
 44.40&$\pm$0.03~\cite{lum1} & 4226.26$\pm0.04$~\cite{ecm1}& 2013 \\\hline
 530.30&$\pm2.39$~\cite{ecm2} & 4235.77$\pm0.04$~\cite{ecm2} & 2017 \\ \hline
 55.59&$\pm0.04$~\cite{lum1} & 4241.66$\pm 0.12$~\cite{ecm1} & 2013  \\ \hline
 538.10&$\pm2.69$~\cite{ecm2}  & $4243.97\pm0.04$~\cite{ecm2} & 2017 \\ \hline
 523.74&$\pm$0.10~\cite{lum1} & 4257.97$\pm0.04$~\cite{ecm1}& 2013  \\ \hline
 301.93&$\pm$0.08~\cite{lum1} & 4257.97$\pm0.04$~\cite{ecm1} & 2013  \\ \hline
 531.10&$\pm3.13$~\cite{ecm2}  & 4266.81$\pm0.04$~\cite{ecm2} & 2017 \\ \hline
 175.70&$\pm0.97$~\cite{ecm2}  & 4277.78$\pm0.11$~\cite{ecm2} & 2017  \\ \hline
 \multicolumn{2}{|c|}{502.4} & 4288.43$\pm0.06$~\cite{ecm2} & 2019  \\ \hline
 \multicolumn{2}{|c|}{6.8 to 18.0~\cite{lumscan}} & 29 energies & 2014 \\ \hline
  \end{tabular}
  
  \label{tab:signalData}
\end{table}

The Beijing Spectrometer (BESIII) detector is described in detail in Ref.~\cite{detector}. A super-conducting solenoid provides a 1.0~T magnetic field. Inside the magnet are the multi-layer drift chamber (MDC) for particle tracking, a CsI (Tl) electromagnetic calorimeter (EMC) to measure the energy of electromagnetic showers, and a time-of-flight system (TOF) using plastic scintillators to help with particle identification. Charged particles with a momentum of 1~GeV/c have a momentum resolution of 0.5\%, and the $dE/dx$ resolution is 6\% for electrons from Bhabha scattering. The EMC measures photons with a resolution of 2.5\% (5\%) at 1 GeV in the barrel (end cap) region. The time resolution of the TOF barrel region is 68 ps, while the end cap has a resolution of 110~ps. The end cap TOF was upgraded in 2015 with multi-gap resistive plate chambers, providing a time resolution of 60~ps~\cite{tof}. 

Simulated data samples produced with a {\sc geant4}-based~\cite{geant4} Monte Carlo (MC) package, which includes the geometric description of the BESIII detector and the detector response, are used to determine detection efficiencies and to estimate backgrounds. The simulation models the beam energy spread and initial state radiation (ISR) in the $e^+e^-$ annihilations with the generator {\sc kkmc}~\cite{kkmc}. The inclusive MC simulation sample includes the production of open charm processes, the ISR production of vector charmonium(-like) states, and the continuum processes incorporated in KKMC \cite{kkmc}. The known decay modes are modeled with {\sc evtgen}~\cite{evtgen} with the branching fractions taken from the Particle Data Group (PDG)~\cite{pdg}, and the remaining unknown charmonium decays are modeled with {\sc lundcharm}~\cite{lund}. Final state radiation~(FSR) from charged final state particles is incorporated using the {\sc photos} package~\cite{photos}.

Signal MC samples are generated for the search channels to estimate the reconstruction efficiency. We assume the E1 transition dominates $e^+e^-\to\gamma X(3872)$, so the angular distribution is given by $1-\frac{1}{3}\cos^2\theta_\gamma$ \cite{Eichten:1974af}, where $\theta_\gamma$ is the helicity angle of the photon. The decays of $X(3872)\to\pi^+\pi^-\chi_{c0}$ and $X(3872)\to\pi^0\pi^0\chi_{c0}$ are generated with a uniform distribution in phase space. The $X(3872)\to\pi^0\chi_{c0}$ decay is generated as a P-wave transition. Uniform distributions in phase space are also used to model both decays of $\chi_{c0}\to\pi^+\pi^-$ and $\chi_{c0}\to K^+K^-$. For the four-body $\chi_{c0}$ decays, we include a uniform phase space component as well as the most common intermediate states through which the $\chi_{c0}$ can decay, with sizes scaled according to the branching fractions measured in the PDG. This means for $\chi_{c0}\to\pi^+\pi^-\pi^+\pi^-$ decays we include both $\chi_{c0}\to\pi^+\pi^-\pi^+\pi^-$ and $\chi_{c0}\to\rho^0\pi^+\pi^-$. The intermediate state included for $\chi_{c0}\to\pi^+\pi^-K^+K^-$ is $\chi_{c0}\to K_1^\pm(1270) K^\mp$, and for $\chi_{c0}\to\pi^+\pi^-\pi^0\pi^0$ it is $\chi_{c0}\to\rho^\pm\pi^\mp\pi^0$. To optimize the selection criteria, we normalize the size of the $X(3872)\to\pi\pi \chi_{c0}$ MC sample by setting $\mathcal{B}(X(3872)\to\pi\pi\chi_{c0})=\mathcal{B}(X(3872)\to\pi^+\pi^-J/\psi)$, while for $X(3872)\to\pi^0\chi_{c0}$, we scale the signal MC sample to match the branching fraction ratio predicted by Ref.~\cite{voloshin} for a four-quark state, \RDOUBLE$=2.97$. 

We also generate background MC samples for the processes $e^+e^-\to\omega\chi_{c0}$~\cite{omegaChi}, $e^+e^-\to\pi\pi\psi(2S)$~\cite{pipiPsi2S}, $e^+e^-\to\pi^0\psi(2S)$, $e^+e^-\to\eta\psi(2S)$, and $e^+e^-\to\gamma_{ISR}\psi(2S)$, since all of these backgrounds could peak at the $\chi_{c0}$ when $\psi(2S)\to\gamma\chi_{c0}$. We generate $e^+e^-\to\pi^0\psi(2S)$ and $e^+e^-\to\eta\psi(2S)$ samples of the same size as $e^+e^-\to\pi^+\pi^-\psi(2S)$ to get a conservative estimation of the size of these backgrounds. The $\chi_{c0}$ decays are the same as in the signal MC samples, and the other particles are allowed to decay inclusively.

To ensure that any potential signal is not due to cross feed from other $X(3872)$ decay channels, we also generate background MC samples for all known $X(3872)$ decays. We simulate $e^+e^-\to\gamma X(3872)$, with the $X(3872)$ state decaying to $\gamma J/\psi$, $\pi^0\chi_{c1}$, $\rho^0J/\psi$, $\omega J/\psi$, and $D^0\overline{D}^{*0}+c.c.$, with all particles decaying inclusively. We also simulate $X(3872)\to\gamma\psi(2S)$ using the central value measured by LHCb ~\cite{gamma}, even though BESIII sets a more stringent upper limit on this decay mode. 

The inclusive MC sample is used to check for other possible backgrounds, and no peaking backgrounds are found in the $X(3872)$ signal region.

\section{Event Selection}

All the final state particles are required to be reconstructed in the detector. Charged tracks detected in the MDC are required to be within a polar angle ($\theta$) range of $|\rm{cos\theta}|<0.93$, where $\theta$ is defined with respect to the $z$-axis. The distance of closest approach to the interaction point (IP) must be less than 10\,cm along the $z$-axis, $|V_{z}|$, and less than 1\,cm in the transverse plane, $|V_{xy}|$. No particle identification is used.

Photon candidates are identified using showers in the EMC.  The deposited energy of each shower must be more than 25~MeV in the barrel region ($|\cos \theta|< 0.80$) and more than 50~MeV in the end cap region ($0.86 <|\cos \theta|< 0.92$).
To exclude showers that originate from charged tracks, the angle between the position of each shower in the EMC and the closest extrapolated charged track must be greater than 10 degrees. To suppress electronic noise and showers unrelated to the event, the difference between the EMC time and the event start time is required to be within (0, 700)\,ns.

We perform a (4+n)C kinematic fit, where 4 constraints conserve the total four-momentum and the rest constrain the masses of the n $\pi^0$s in the final state. The $\chi^2/DOF$, where $DOF$ is the number of degrees of freedom in the kinematic fit, is optimized for each final state, and is the selection criteria that removes the most background. We reconstruct $\chi_{c0}$ candidates between 3.2 and 3.7 GeV$/c^2$ so we can use $\chi_{c0}$ sidebands to estimate the shape and size of the non-peaking backgrounds. We fit $X(3872)$ candidates between 3.75 and 4.0 GeV/$c^2$ after selecting $\chi_{c0}$ candidates in a 50 MeV/$c^2$ window centered on the PDG mass of the $\chi_{c0}$ resonance~\cite{pdg}. We only use data in the region $4.15 < E_{\rm CM} < 4.3$ GeV for our fits, since that is where the production cross section is the largest~\cite{besX}.

For the normalization channel, we use the same event selection as Ref.~\cite{ryan}. We separate electrons and muons based on the energy deposited in the EMC divided by the track momentum ($E/p$). Electrons are required to have $E/p>0.85$, while muons must have $E/p<0.25$. Additionally, we require the kinematic fit has $\chi^2/DOF<10$. The $J/\psi$ candidate is selected by using a 40 MeV$/c^2$ window in the $\ell^+\ell^-$ invariant mass distribution centered on the PDG mass of the $J/\psi$ resonance. To suppress radiative Bhabha events, the opening angle between the pions is required to satisfy $\cos\theta_{\pi\pi}<0.98$. This background is further suppressed by requiring the angle between any charged track and photon satisfies $\cos\theta_{\gamma tk}<0.98$. There are additional backgrounds from $e^+e^-\to\eta J/\psi$ and $e^+e^-\to\eta'J/\psi$ for the normalization channel, and they are suppressed by requiring $M(\gamma\pi^+\pi^-)>0.6$ GeV/$c^{2}$ and $| M(\gamma\pi^+\pi^-)-M(\eta')|>0.02$ GeV/$c^2$, where $M(\eta')$ is the PDG mass of the $\eta'$ resonance.  

\begin{figure*}
  \centering
  \includegraphics[scale=0.3]{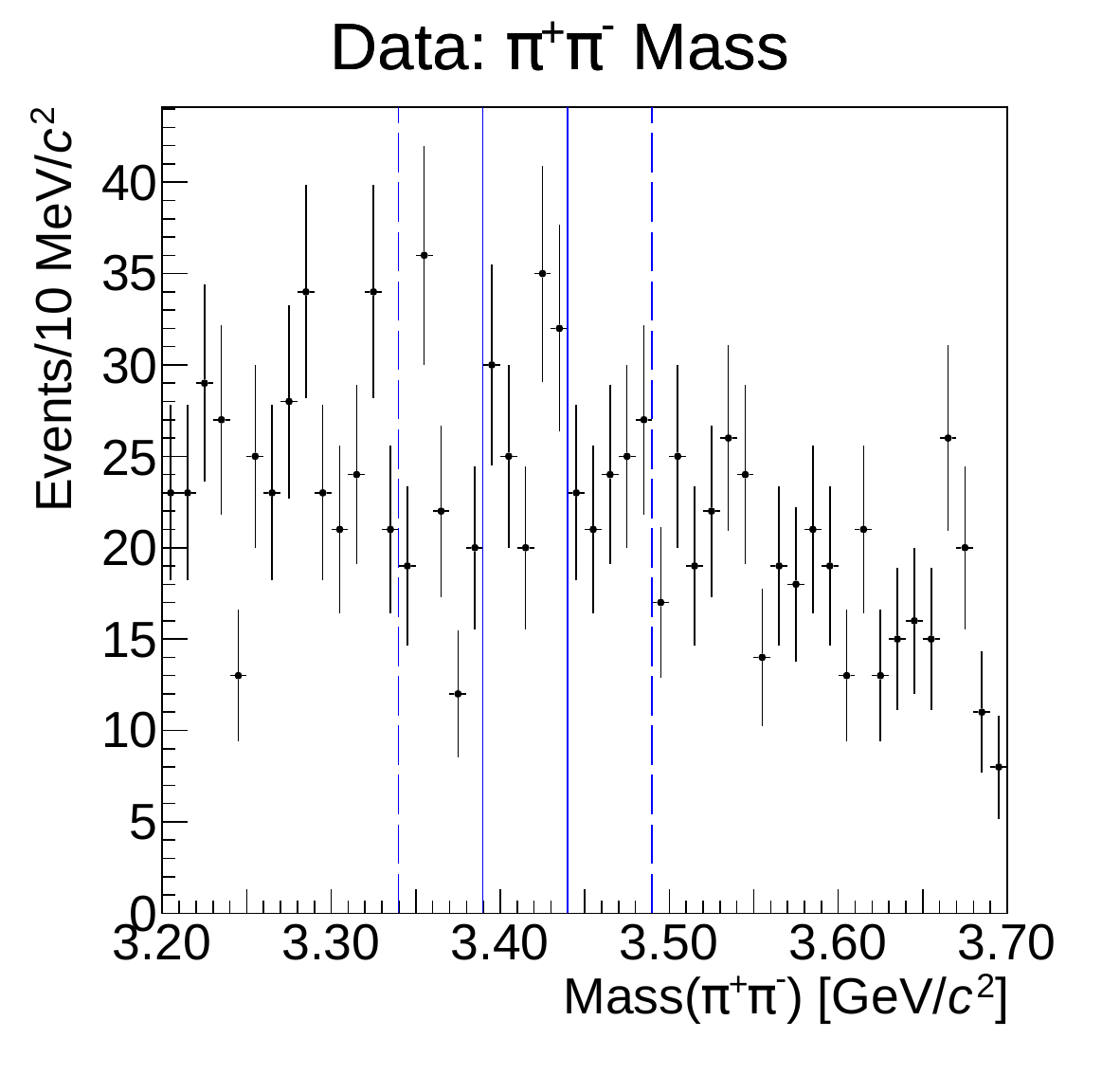}\includegraphics[scale=0.3]{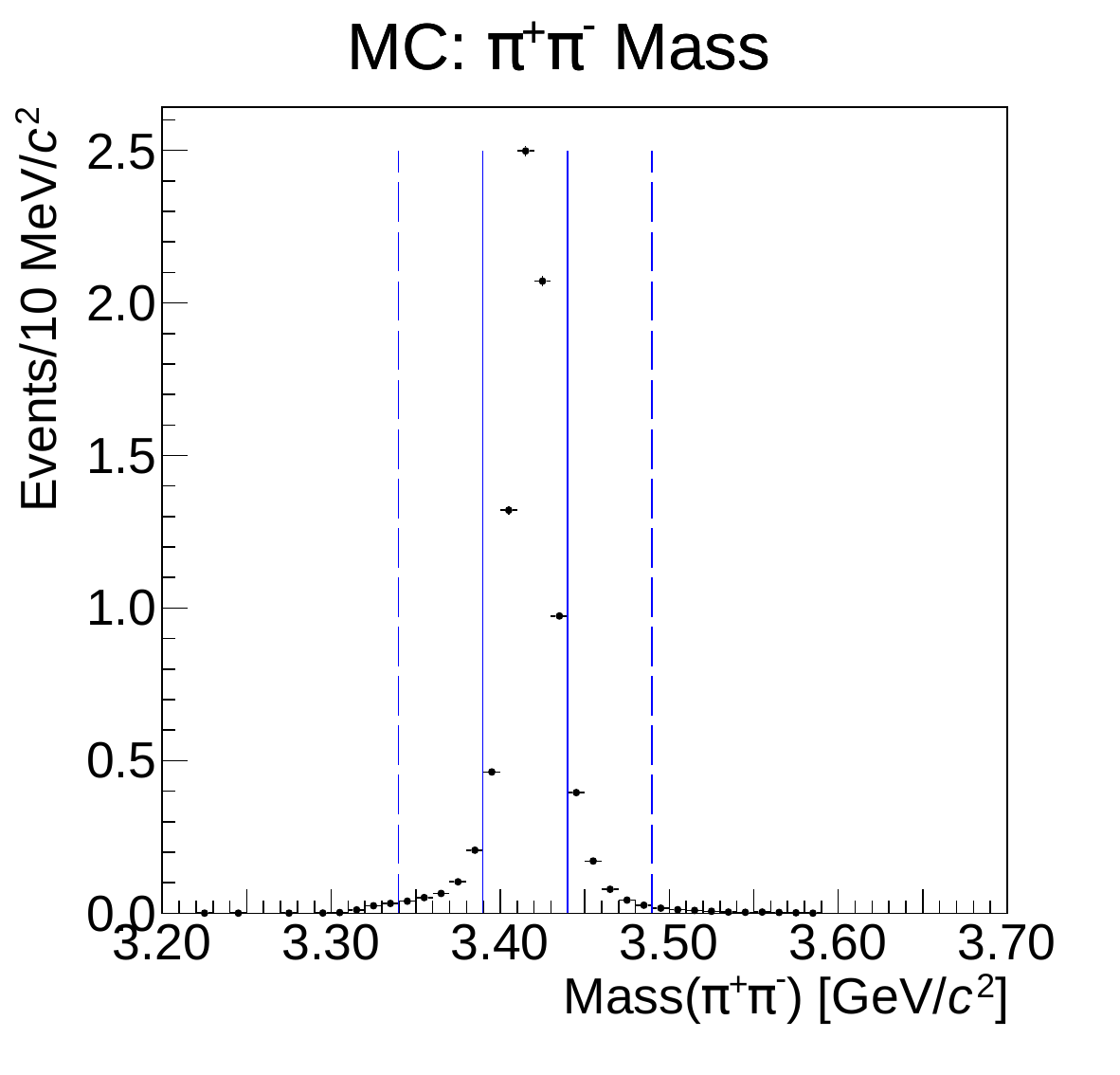}\\\includegraphics[scale=0.3]{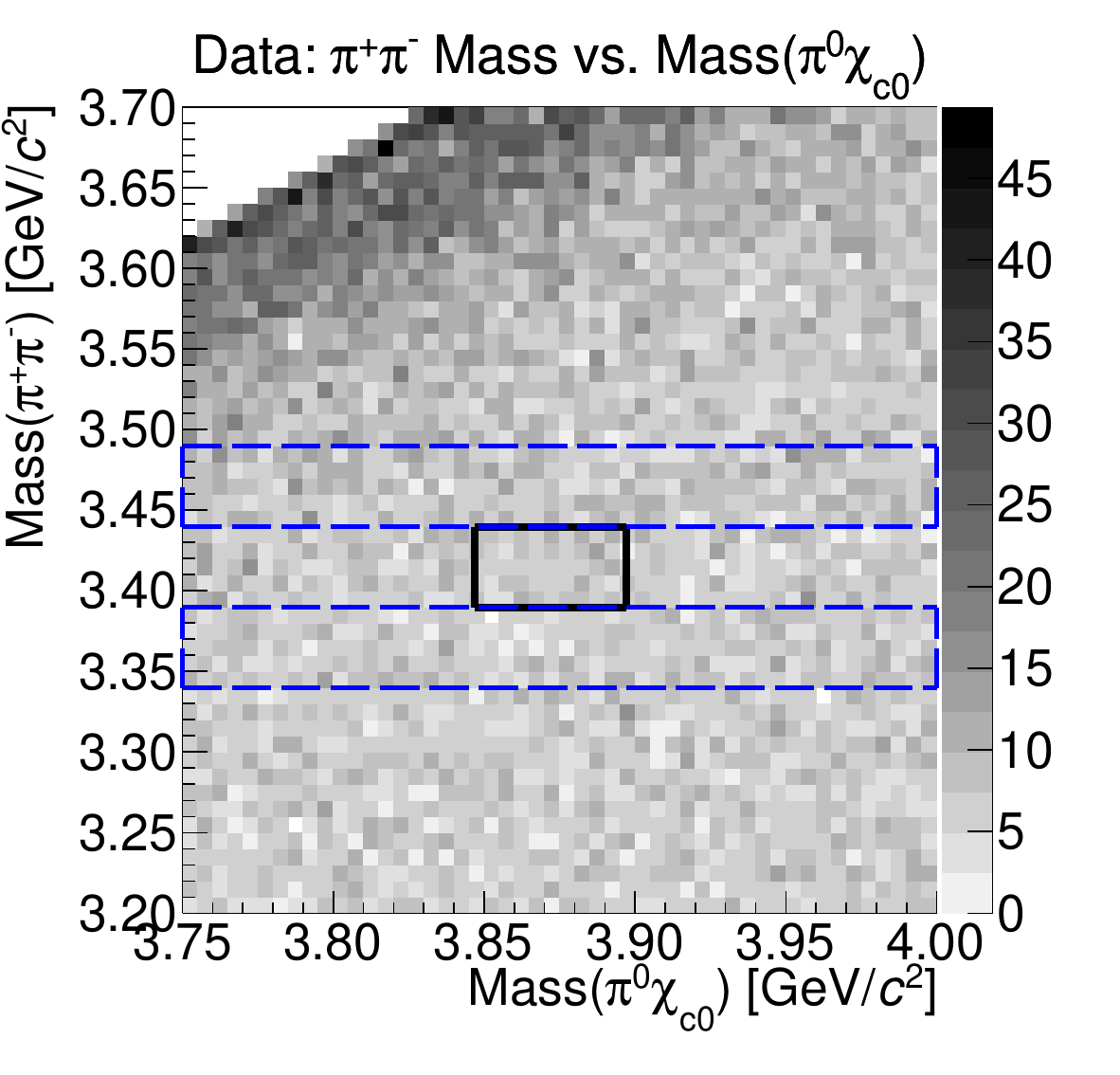}\includegraphics[scale=0.3]{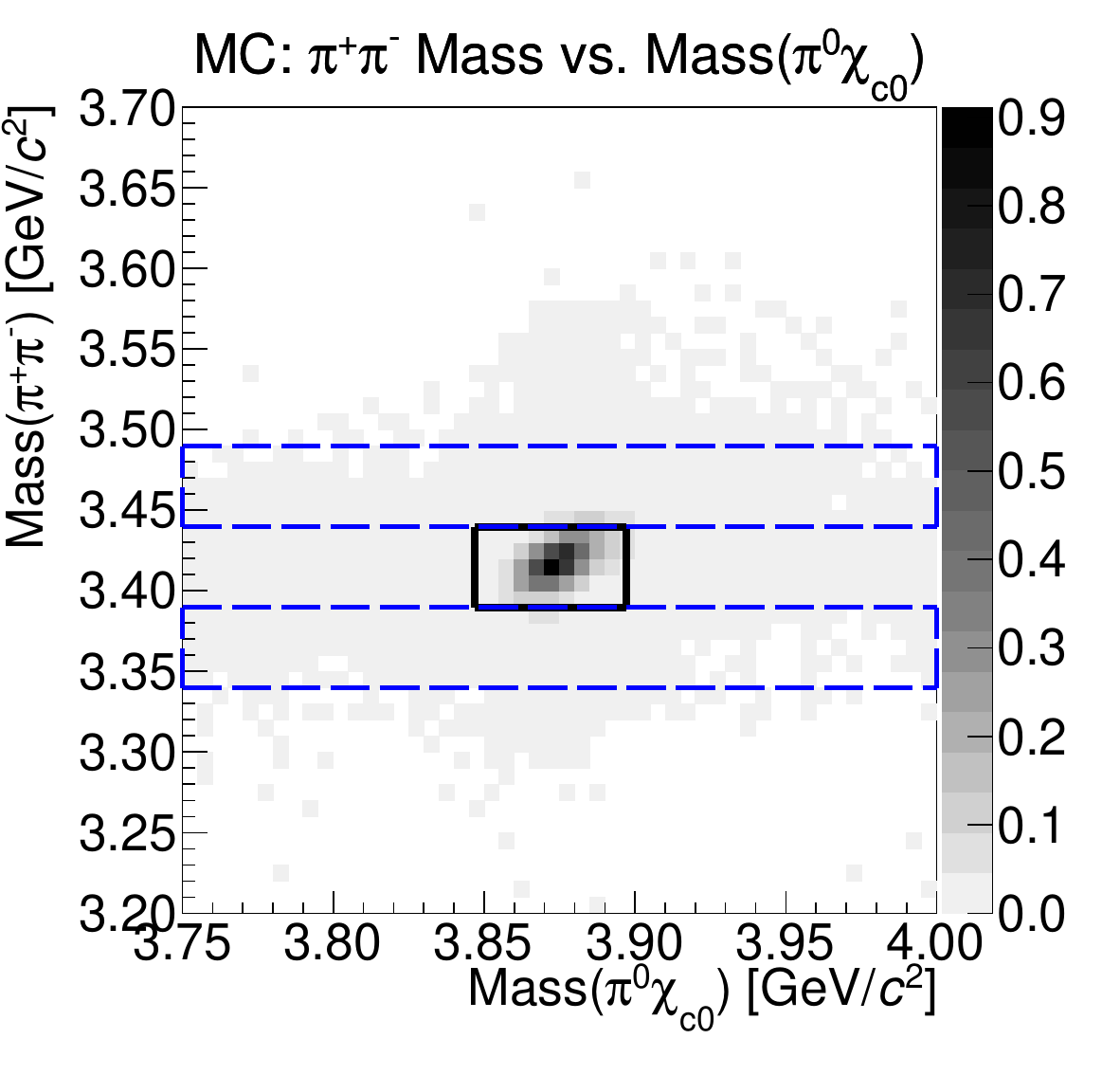}
  \caption{Plots showing the signal and sideband regions in data (left) and signal MC (right) before selection criteria optimization. The top row shows the $\chi_{c0}$ candidate mass projections, where the solid lines denote the signal region and the two sideband regions are the areas between the solid and dashed lines. The bottom row shows two-dimensional plots of the $\chi_{c0}$ candidate masses versus the $X(3872)$ candidate masses, where the z-axis is the number of events in each bin. The central (solid) box is the signal region, and the long (dashed) rectangles above and below the $\chi_{c0}$ mass are the sidebands used to estimate non-peaking $\chi_{c0}$ backgrounds.}
  \label{fig:2d}
\end{figure*}

In order to refine the selection criteria for the search channels, we optimize the figure of merit $\textrm{FOM}=\frac{S}{\sqrt{S+B_1+B_2}}$. The signal $S$ is from a 50 MeV/$c^2$ box centered on the PDG masses of the $X(3872)$ and $\chi_{c0}$ states in the signal MC samples, scaled according to Ref.~\cite{voloshin}. The background $B_1$ is the estimated number of background events in the signal region of the $X(3872)$ using $\chi_{c0}$ sidebands that are 50 MeV/$c^2$ wide in the $\chi_{c0}$ candidate masses on both the lower and higher mass sides of the $\chi_{c0}$ signal region in data and extend from 3.75 to 4.0 GeV$/c^2$ in the $X(3872)$ candidate masses. The background $B_2$ is the estimated number of peaking $\chi_{c0}$ background events, which is determined using the peaking $\chi_{c0}$ background MC samples, which are scaled according to previous measurements at BESIII~\cite{omegaChi, pipiPsi2S}. A plot showing the signal region and sideband regions for $X(3872)\to\pi^0\chi_{c0}$ decays with $\chi_{c0}\to\pi^+\pi^-$ is shown in Figure~\ref{fig:2d}. The data plots in the left column show that the $\chi_{c0}$ sidebands match the background in the signal region well, and the signal MC plots in the right column show the selection window includes the majority of our signal. Note that the upper right plot shows that part of the signal MC falls in the sideband region. This $\chi_{c0}$ selection window was optimized by maximizing the FOM, and includes approximately 87\% to 90\% of events for most final states. Since there are two $\chi_{c0}$ sidebands that are each five times as wide as the signal region in the $X(3872)$ invariant mass distribution, this means we scale the sidebands down by a factor of ten for the FOM calculation. These wide sidebands in $X(3872)$ candidate masses are used to increase the data sample size for the FOM calculation. For final states that have multiple possible $\chi_{c0}$ combinations, for instance, $X(3872)\to\pi^0\chi_{c0}$ with $\chi_{c0}\to\pi^+\pi^-\pi^0\pi^0$, an event is in the signal region if at least one combination of the $\pi^+\pi^-\pi^0\pi^0$ invariant mass is in the signal region, and a sideband event if at least one combination is in the sidebands and no combinations fall in the signal region.

In order to use a variable to suppress background, we require the FOM increases by at least 5\%. This is done to simplify the selection criteria as well as to reduce the systematic uncertainties. We optimize the $\chi^2/DOF$ of the kinematic fit for each $\chi_{c0}$ decay mode, which is the requirement that removes most of the background. In some cases, the FOM forms a broad plateau for different $\chi^2/DOF$ values, so we choose the loosest requirement that gives an FOM value within 1\% of the maximum. By doing this, we have essentially the same FOM value, but a much larger reconstruction efficiency. The $\chi^2/DOF$ requirements for each final state range from 2.25 to 8.0, and the exact values are shown in Table~\ref{tab:optCutpi0}. We also optimize $E/p$ to separate pions from electrons and positrons. This criteria is used for two final states, $X(3872)\to\pi^0\chi_{c0}$ with $\chi_{c0}\to\pi^+\pi^-$ requires $E/p<0.95$, and $X(3872)\to\pi^+\pi^-\chi_{c0}$ with $\chi_{c0}\to\pi^+\pi^-$ requires $E/p<0.85$. This requirement does not improve the FOM more than 5\% for the other final states.

\begin{table}[ht]
  \caption{Optimized $\chi^2/DOF$ values for all three $X(3872)$ decays.}
  \centering
\begin{tabular}{ |c|c|c|c| }
\hline
 Decay & $X\to\pi^0\chi_{c0}$ & $X\to\pi^+\pi^-\chi_{c0}$ & $X\to\pi^0\pi^0\chi_{c0}$ \\ \hline
 $\chi_{c0}\to\pi^+ \pi^-$ & 4.0 & 6.0 & 5.25 \\ \hline
 $\chi_{c0}\to K^+ K^-$ & 3.75  & 4.25 & 2.5 \\ \hline
 $\chi_{c0}\to\pi^+\pi^-\pi^+\pi^-$ & 3.0 & 8.0 & 3.75 \\ \hline
$\chi_{c0}\to\pi^+\pi^- K^+ K^-$ & 3.75 & 6.5 & 3.75 \\ \hline
 $\chi_{c0}\to\pi^+\pi^-\pi^0 \pi^0$ & 2.25 & 3.0 & 2.5 \\ \hline
\end{tabular}

\label{tab:optCutpi0}
\end{table}

Several additional variables are investigated to reduce the background, but none of them increase the FOM enough to be included. These include a $\psi(2S)$ veto on the $\gamma\chi_{c0}$ invariant mass, an $\omega$ veto on the $\gamma\pi^0$ system, and a veto on $D\to\pi K$. The background $e^+e^-\to\gamma D^* \bar{D}$ with $D^*\to \gamma D$ with $D\to \pi^+\pi^0 K^-$ and $\bar{D}\to K^+\pi^-$ gives the same final state as the search channel $X(3872)\to\pi^0\chi_{c0}$ with $\chi_{c0}\to\pi^+\pi^-K^+K^-$, but inclusive MC samples show this background does not cause any peaking backgrounds for $4.15<E_{\rm CM}<4.3$ GeV.

There is a small amount of peaking background at the $X(3872)$ mass for several decay modes. Only two search final states have at least one predicted $X(3872)$ background event. The $X(3872)\to\pi^0\pi^0\chi_{c0}$ channel with $\chi_{c0}\to\pi^+\pi^-K^+K^-$ has a background from $X(3872)\to D^{*0}\bar{D}^{0}$ with $D^{*0}\to\pi^0D^0$ and $D^0\to K^-\pi^+\pi^0$ and $\bar{D}^0\to K^+\pi^-$, where the charge conjugated mode is also implied. There will be $1.2\pm0.1$ events due to $X(3872)\to D^0\bar{D}^{*0}+c.c.$ for this search channel. The $X(3872)\to\pi^+\pi^-\chi_{c0}$ mode with $\chi_{c0}\to\pi^+\pi^-\pi^0\pi^0$ has a predicted rate of $1.0\pm0.1$ background events from $X(3872)\to\omega J/\psi$, with $\omega\to\pi^+\pi^-\pi^0$ and $J/\psi\to\pi^+\pi^-\pi^0$ decays. Since the background levels are low, we do not veto them since they give a more conservative upper limit and it keeps the selection criteria simple.

After optimizing the selection criteria, there could still be multiple $X(3872)$ candidates due to the different photon combinations used to reconstruct the $\pi^0$ candidates. To eliminate this double counting, we rank the remaining combinations by their $\chi^2/DOF$ of the kinematic fit and choose the best combination. This leaves at most one $X(3872)$ combination per event, so there is no double counting when we fit the $X(3872)$ mass spectrum. We measure the average number of combinations per event in the full range of $X(3872)$ and $\chi_{c0}$ candidate masses before selecting the best combination, and find that for all $\chi_{c0}$ decays except $\chi_{c0}\to\pi^+\pi^-K^+K^-$ it ranges from 1.0 to 1.2 in both data and signal MC. For the $\chi_{c0}\to\pi^+\pi^-K^+K^-$ mode, there can be multiple combinations from swapping a pion and kaon during the reconstruction. In this case, the average number of combinations before selecting the best combination varies from 1.5 to 1.8 in data and signal MC simulation. Signal Monte Carlo studies show that the correct $X(3872)$ candidate is selected approximately 99\% of the time for all final states.

\section{Fitting}

The production mechanism for the search and normalization channels are the same, so we are able to combine data from different energies in our measurements. Unbinned extended maximum likelihood fits are performed to the $X(3872)$ candidate mass spectrum in the region 3.75 to 4.0 GeV/$c^2$ for both the signal and normalization channels. 

The $X(3872)$ signal shape for the normalization channel is well described by a Voigtian function, the convolution of a Breit-Wigner and a Gaussian function. The measured $X(3872)$ width of 0.96 MeV from Ref.~\cite{narrow1} is used to fix the internal width of the Breit-Wigner. We fit the signal MC sample to fix the mass and resolution parameters of the Gaussian function. This fit is also used to determine the reconstruction efficiency, which is calculated by integrating the fit function and dividing by the number of events that were generated. To combine the efficiencies at different energy points, we perform a weighted average, where the weights are the luminosity times cross section for each $E_{CM}$. The default cross section used is the $\sigma(e^+e^-\to\gamma X(3872))$ measured in Ref.~\cite{besX}. The total fit function for the normalization channel is the sum of the Voigtian signal function and a first order polynomial function to describe the background.

Each $X(3872)$ search channel includes five $\chi_{c0}$ decay modes, so we perform a simultaneous fit to all five final states. The Voigtian signal shapes and the reconstruction efficiencies are determined using the same method as the normalization channel. We scale the relative sizes of the signal yield for each $\chi_{c0}$ decay mode. To do so, we use scaling factors $w_i=\epsilon_i\mathcal{B}_i(\chi_{c0})$, where $\epsilon_i$ is the reconstruction efficiency and $\mathcal{B}_i(\chi_{c0})$ is the PDG branching fraction for a specific $\chi_{c0}$ decay mode $i$. Using these scales, we define

\begin{equation*}
 N_{\textrm{tot},i} \equiv\frac{N_i}{w_i},
\end{equation*}
where $N_i$ is the signal yield for a specific $\chi_{c0}$ decay mode $i$.
We then constrain the $N_{\textrm{tot},i}$ to be the same for all five $\chi_{c0}$ decay modes, so we get a single yield value $N_{\textrm{tot}}$ from the fit.

The $X(3872)$ search channels all have a background process with the same final state, including a $\chi_{c0}$. For $e^+e^-\to\gamma X(3872)$ with $X(3872)\to\pi^0\chi_{c0}$, this is $e^+e^-\to\omega\chi_{c0}$ with $\omega\to\gamma\pi^0$, and for $e^+e^-\to\gamma X(3872)$ with $X(3872)\to\pi\pi\chi_{c0}$, it is $e^+e^-\to\pi\pi\psi(2S)$, with $\psi(2S)\to\gamma\chi_{c0}$. All of these processes produce an asymmetric background shape in the $X(3872)$ candidate mass spectrum. To account for this background, we include a histogram of these background processes in the fit. The size and shape of this histogram are fixed based on MC samples that were generated using previously measured cross sections at BESIII. The non-$\chi_{c0}$ peaking backgrounds are described using a first order polynomial function for $X(3872)\to\pi^0\chi_{c0}$ and second order polynomial functions for $X(3872)\to\pi\pi\chi_{c0}$. In order to more accurately determine the shape of the non-$\chi_{c0}$ peaking backgrounds, we constrain the shape of the background polynomial functions to be the same in the $\chi_{c0}$ signal and sideband regions in data. To ensure no signal contaminates the sideband fit, we exclude the $X(3872)$ signal region in the fit to the sideband regions. We include a floating scale factor between the polynomial function in the signal and sideband regions. The total fit functions for the search channels are the sum of the polynomial function, the peaking $\chi_{c0}$ background MC histogram, and the Voigtian signal function.

To measure the ratio of branching fractions, we use the formula

\begin{equation*}
  \frac{\mathcal{B}(X(3872)\to\pi^0\chi_{c0})}{\mathcal{B}(X(3872)\to\pi^+\pi^-J/\psi)}=\frac{N_{\textrm{tot}}}{\mathcal{B}(\pi^0)}\frac{\epsilon_{\pi\pi J/\psi}\mathcal{B}(J/\psi)}{N_{\pi\pi J/\psi}},
\end{equation*}
where $N_{\pi\pi J/\psi}$ and $\epsilon_{\pi\pi J/\psi}$ are the number of events and reconstruction efficiency for $X(3872)\to\pi^+\pi^-J/\psi$, respectively. The branching fraction for $\pi^0\to\gamma\gamma$ is denoted by $\mathcal{B}(\pi^0)$, and the branching fraction for $J/\psi\to\ell^+\ell^-$ ($\ell=e$ or $\mu$) is $\mathcal{B}(J/\psi)$, which are taken from the PDG. Note that the production cross section, ISR correction factors, and the integrated luminosity are canceled in the ratio, and the ratio of branching fractions is only sensitive to the ratio of efficiencies.

To calculate the significance of the signal, we perform a fit with the signal yield floating, as well as a fit with the signal yield fixed to zero. We then use the likelihood ratio test to determine the statistical significance of the fit. The systematic uncertainty due to the fitting model is determined using 648 alternative fit models, described in detail in Section V F.

\subsection{Normalization Channel}

\begin{figure}
  \includegraphics[scale=0.4]{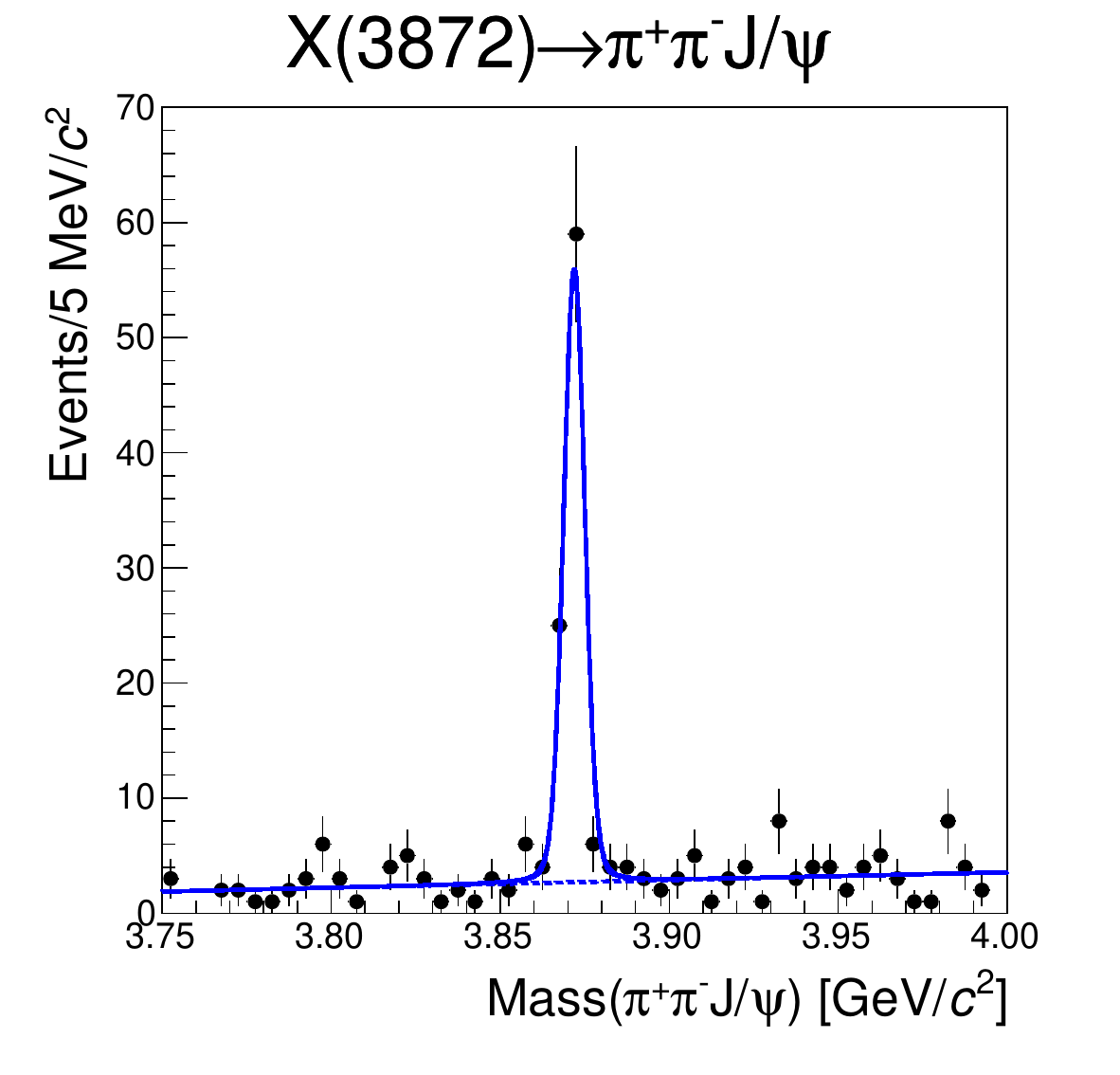}
  \caption{Fit result to the $\pi^+\pi^-J/\psi$ mass spectrum with a first order polynomial function to describe the background (dashed line) and a Voigtian as a signal function (solid line). There is a clear signal for the $X(3872)$ state.}
  \label{fig:norm}
\end{figure}

The results of the fit to the normalization channel are shown in Figure~\ref{fig:norm}. There is a clear signal for the $X(3872)$ state. The reconstruction efficiency is \NORMEFFUNC, and the fit yield is \NORMNUMBER. These are both consistent with Ref.~\cite{ryan}'s measured efficiency of $32.3\%$ and yield of $84.1^{+10.1}_{-9.4}$, where the uncertainties on the yields are just the statistical uncertainties from the fits. Note this analysis includes two more data points not included in Ref.~\cite{ryan}, which is why the yield is larger here.

\subsection{$X(3872)\to\pi^0\chi_{c0}$}

\begin{figure*}
  \includegraphics[scale=0.26]{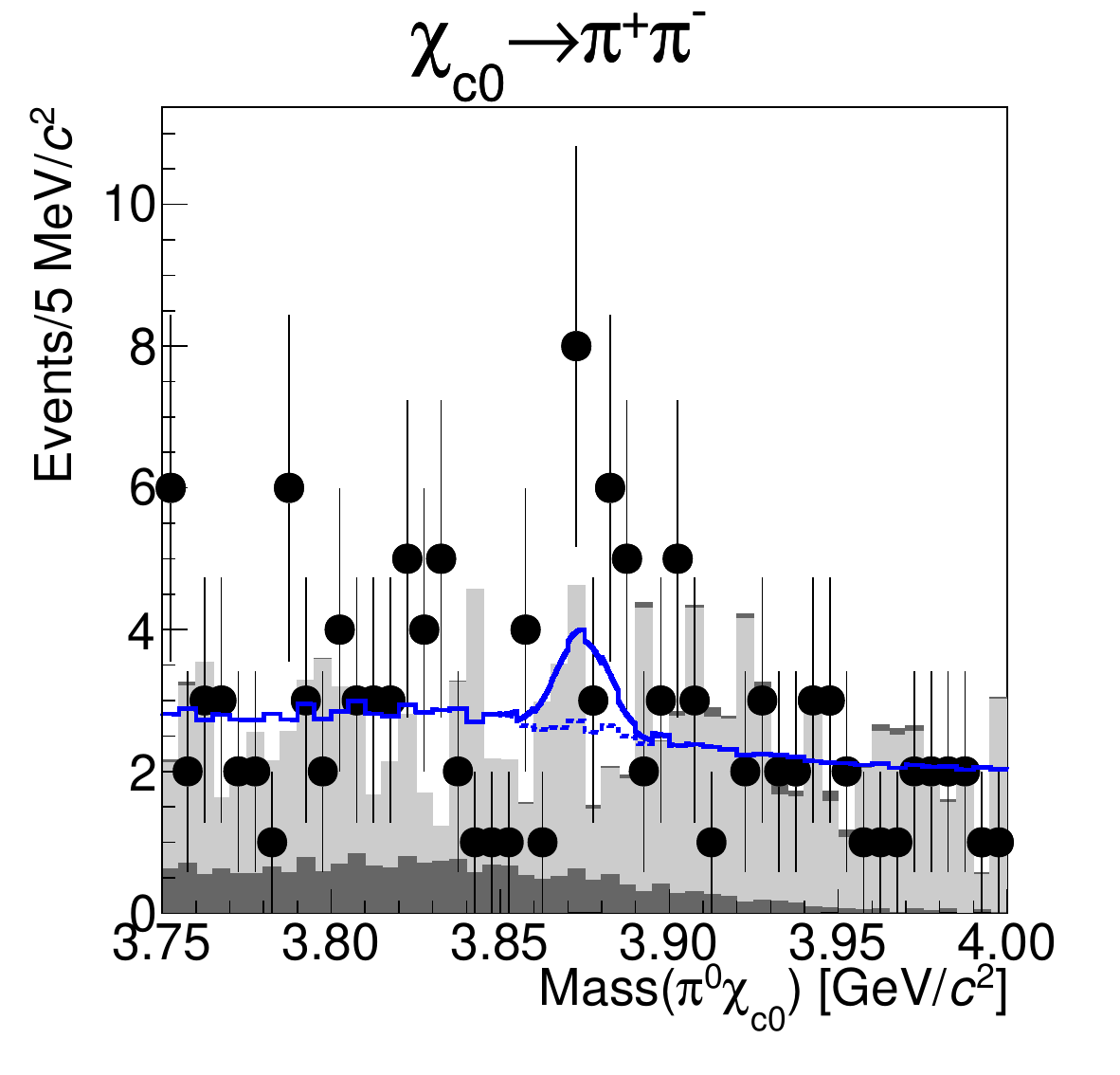}\includegraphics[scale=0.26]{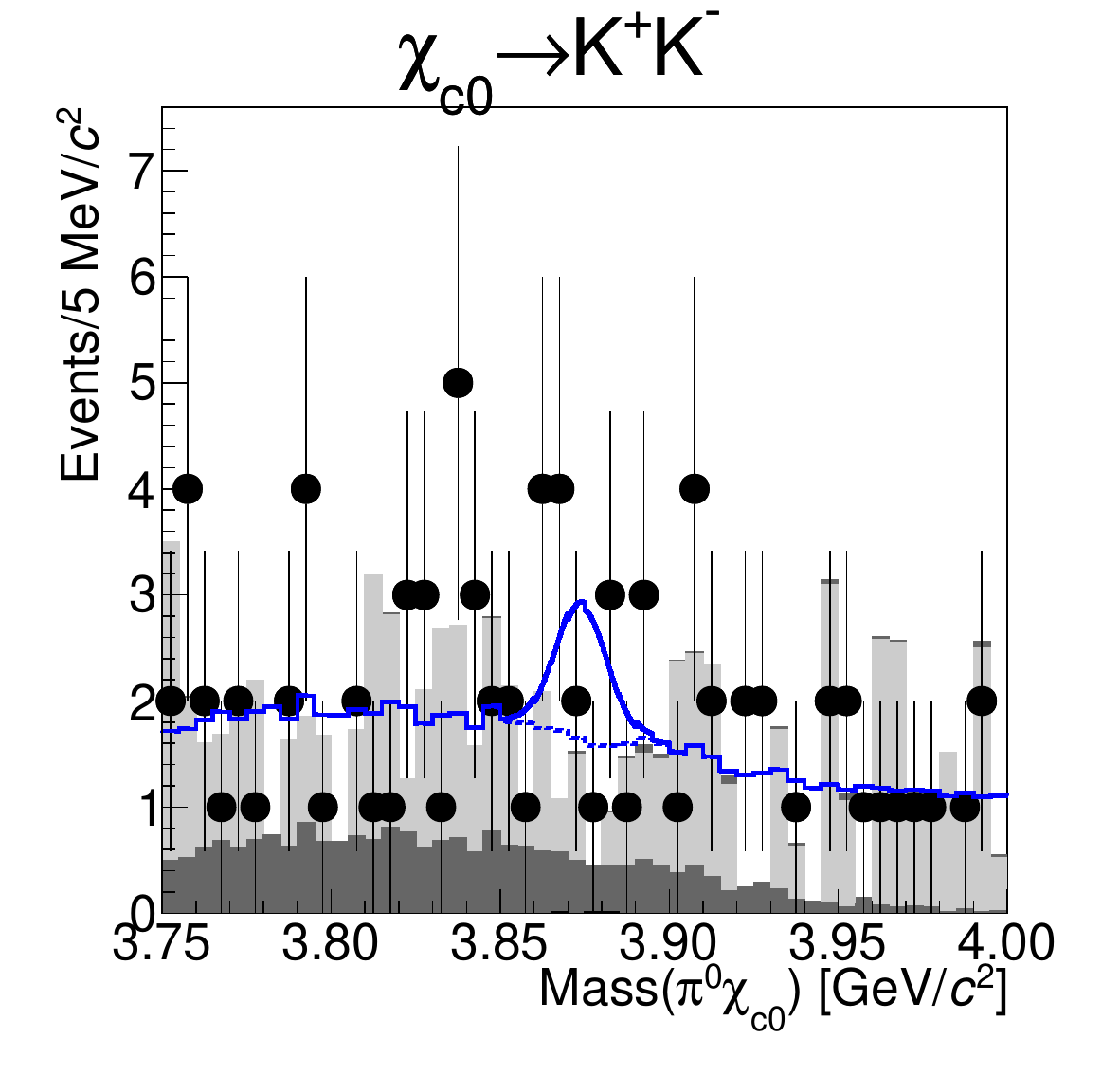} \includegraphics[scale=0.26]{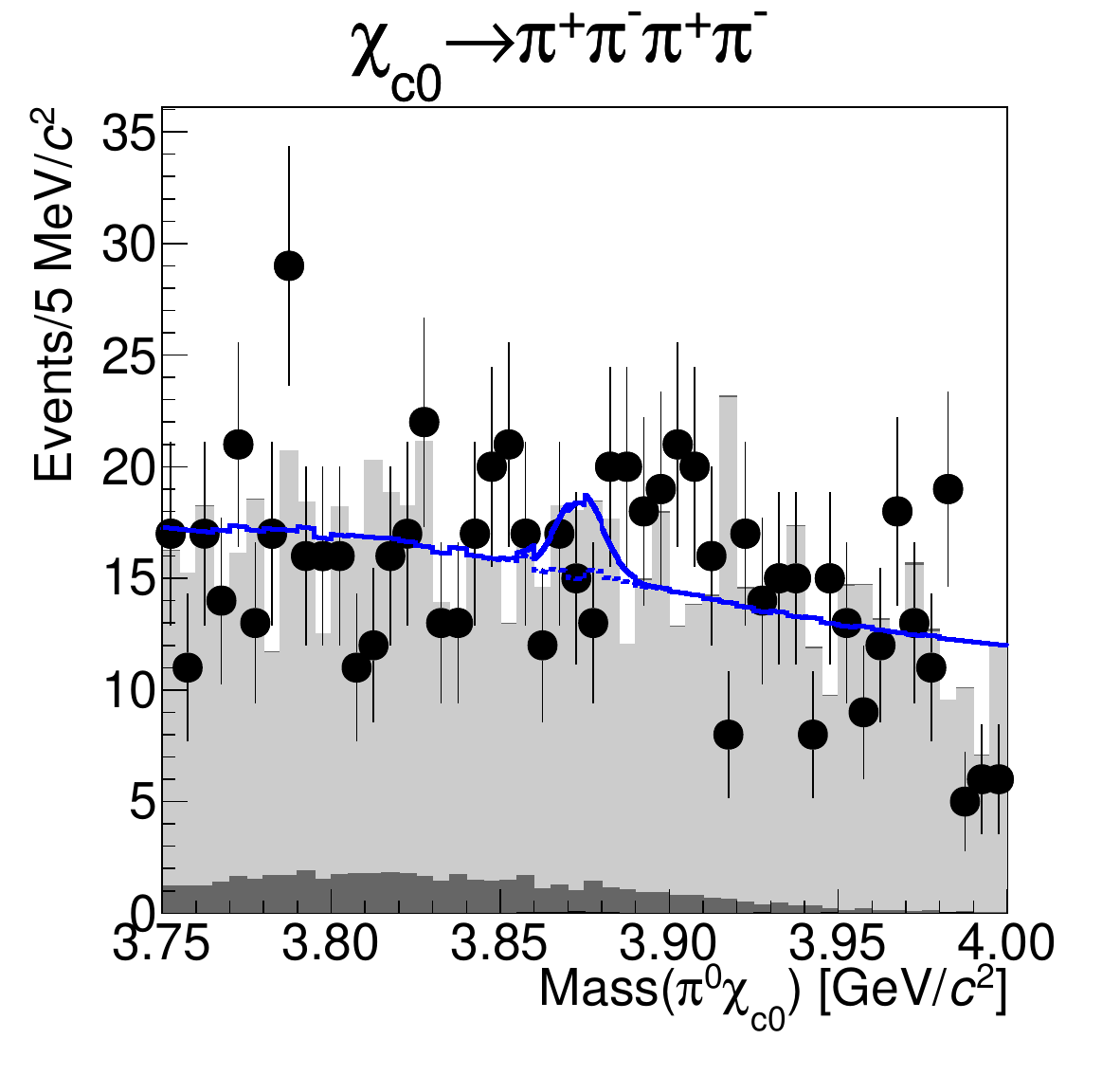}\\\includegraphics[scale=0.26]{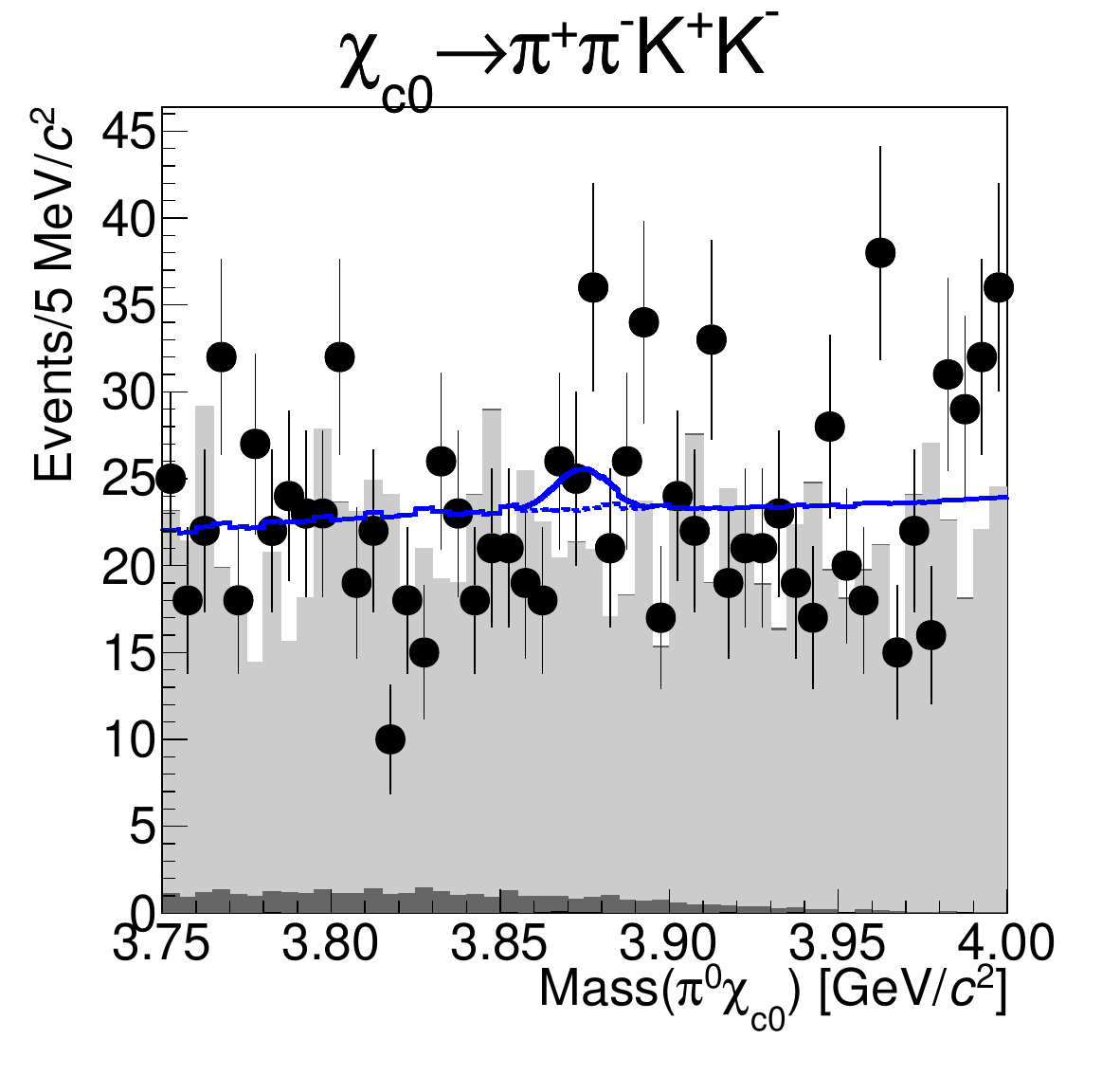}\includegraphics[scale=0.26]{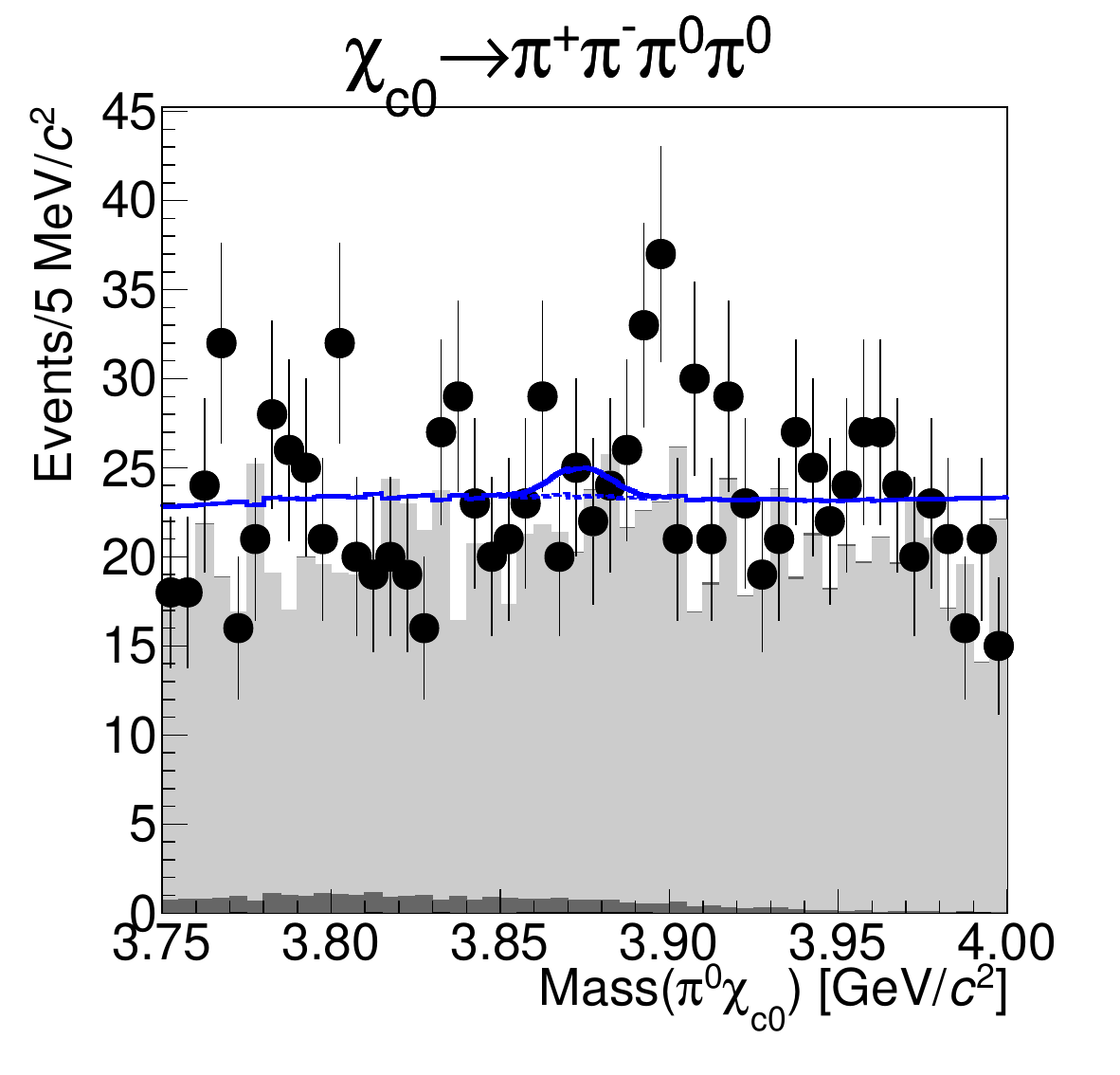}
  \caption{Simultaneous fit to the $\pi^0\chi_{c0}$ mass spectrum for $4.15 < E_{\rm CM} < 4.3$ GeV. The points are data in the signal region, the light gray histogram is the background estimate from $\chi_{c0}$ sidebands in data, and the dark gray histogram is the peaking $\chi_{c0}$ background estimated from MC simulated with $e^+e^-\to\omega\chi_{c0}$. The total background shape is the sum of the $e^+e^-\to\omega\chi_{c0}$ MC shape plus a first order polynomial function. The solid lines show the fit with a signal component, while the dashed lines are the background contributions. The combination of $e^+e^-\to\omega\chi_{c0}$ contribution and $\chi_{c0}$ sidebands describes the size of the backgrounds well.}
  \label{fig:nom}
\end{figure*}

\begin{figure*}
  \includegraphics[scale=0.3]{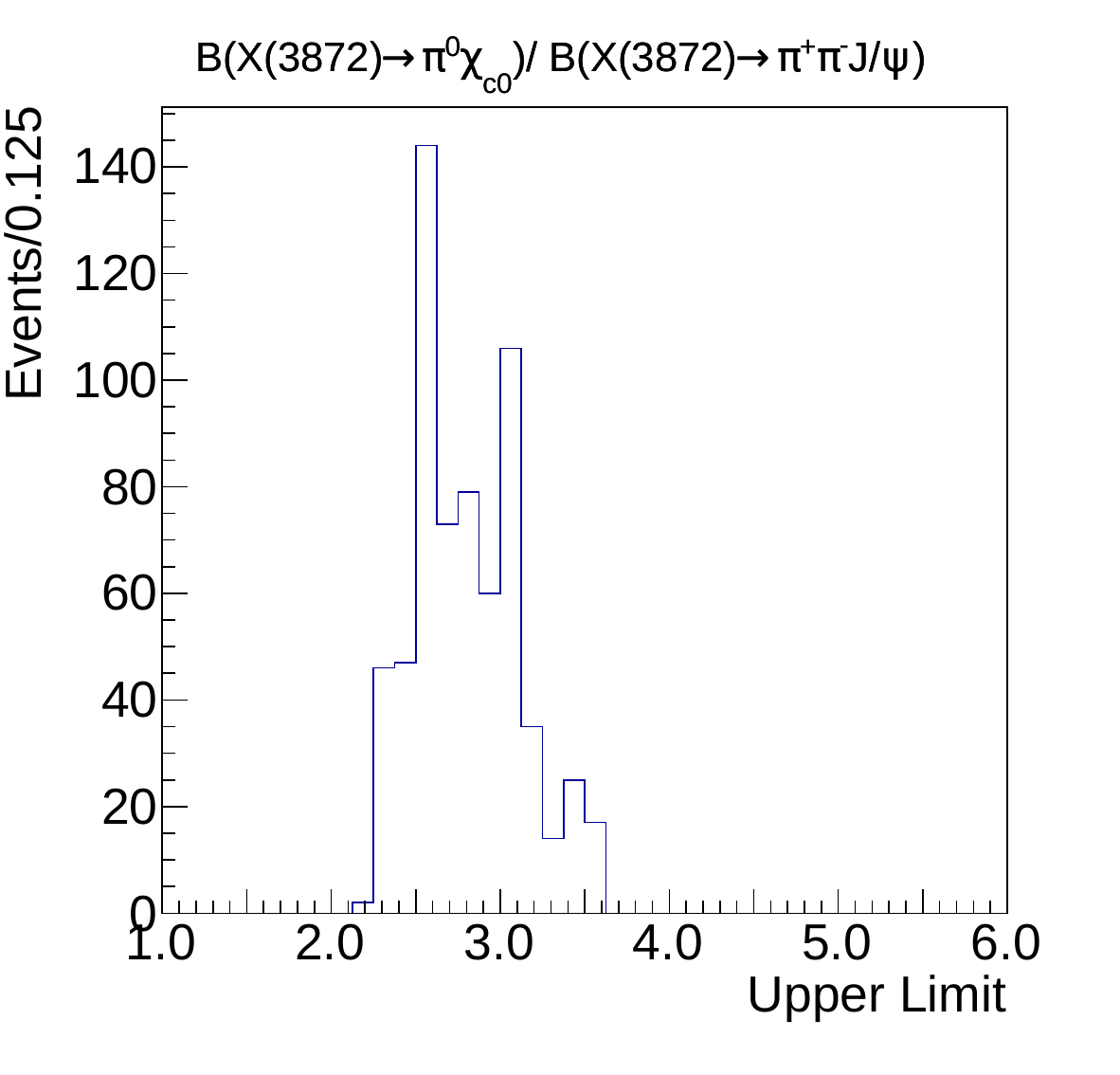}\includegraphics[scale=0.3]{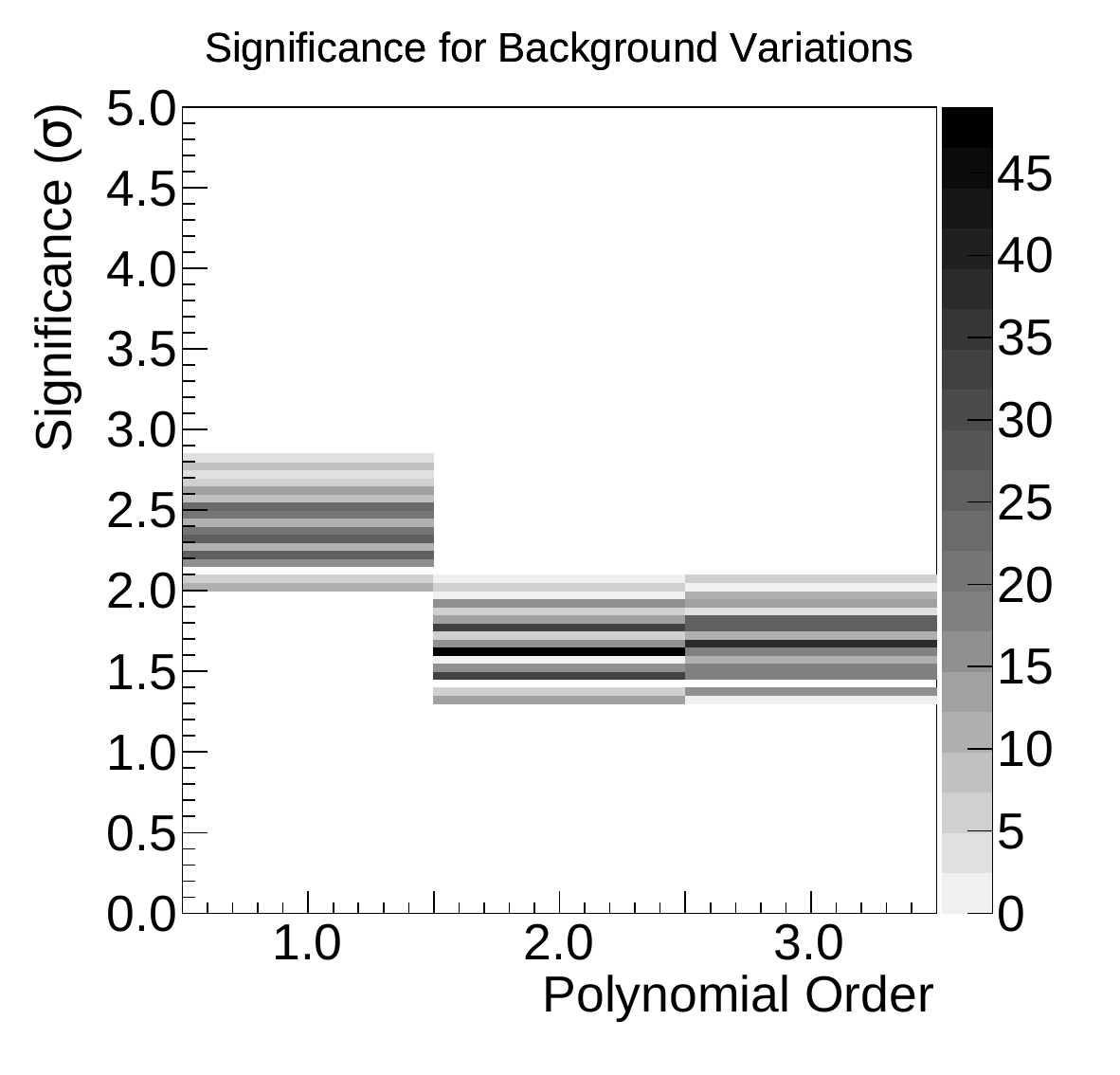}
  \caption{(Left) The distribution of upper limits on \RMAIN~from 648 alternative fits. (Right) Distribution of the significance (y-axis) vs the order of the background polynomial function used (x-axis) for the fit to the $\pi^0\chi_{c0}$ mass spectrum. Here we find the first order polynomial function gives a larger significance than both the second and third order polynomial functions.}
  \label{fig:signif}
\end{figure*}

The fit result to the $\pi^0\chi_{c0}$ mass spectrum is shown in Figure~\ref{fig:nom}. There is no obvious signal for any of the reconstructed $\chi_{c0}$ final states. The four body $\chi_{c0}$ decays have larger background levels than the two body decays, but input/output checks in Monte Carlo show the average significance does not change when high background final states are added. The efficiencies, scaling factors, and yields for each $\chi_{c0}$ decay mode are shown in Table~\ref{tab:nom}. The signal has a total statistical significance of \SIGNIF. In Section V F, we perform 648 alternative fits with different signal and background models. Figure~\ref{fig:signif} shows that the significance for all these fit variations is always at least \SIGNIFMIN. Figure~\ref{fig:signif} also shows the range of upper limit values measured in all the fit variations. 

\begin{table}
  \caption{Efficiencies, scaling factors, and yields for each decay mode of the $\chi_{c0}$ for $X(3872)\to\pi^0\chi_{c0}$. The uncertainties on the yields are statistical only. The scaling factors show the relative contribution of each $X(3872)$ decay in the simultaneous fit. }
  \begin{tabular}{|c|c|c|c|}
\hline
Decay  &  Efficiency  &  Scale ($w_i$)  &  Signal Yield  \\
\hline
$\chi_{c0}\to\pi^+\pi^-$  &  23.4\%  &  0.00133  &  $5.1\pm2.4$  \\
\hline
$\chi_{c0}\to K^+ K^-$  &  21.6\%  &  0.00131  &  $5.0\pm2.3$  \\
\hline
$\chi_{c0}\to \pi^+ \pi^-\pi^+\pi^-$  &  13.5\%  &  0.00315  &  $12.1\pm5.6$  \\
\hline
$\chi_{c0}\to \pi^+\pi^- K^+ K^-$  &  12.9\%  &  0.00233  &  $9.0\pm4.1$  \\
\hline
$\chi_{c0}\to \pi^+\pi^- \pi^0\pi^0$  &  5.83\%  &  0.00188  &  $7.2\pm3.3$  \\
\hline
\end{tabular}
  \label{tab:nom}
\end{table}

\subsection{$X(3872)\to\pi^+\pi^-\chi_{c0}$}

\begin{figure*}
  \centering
  \includegraphics[scale=0.26]{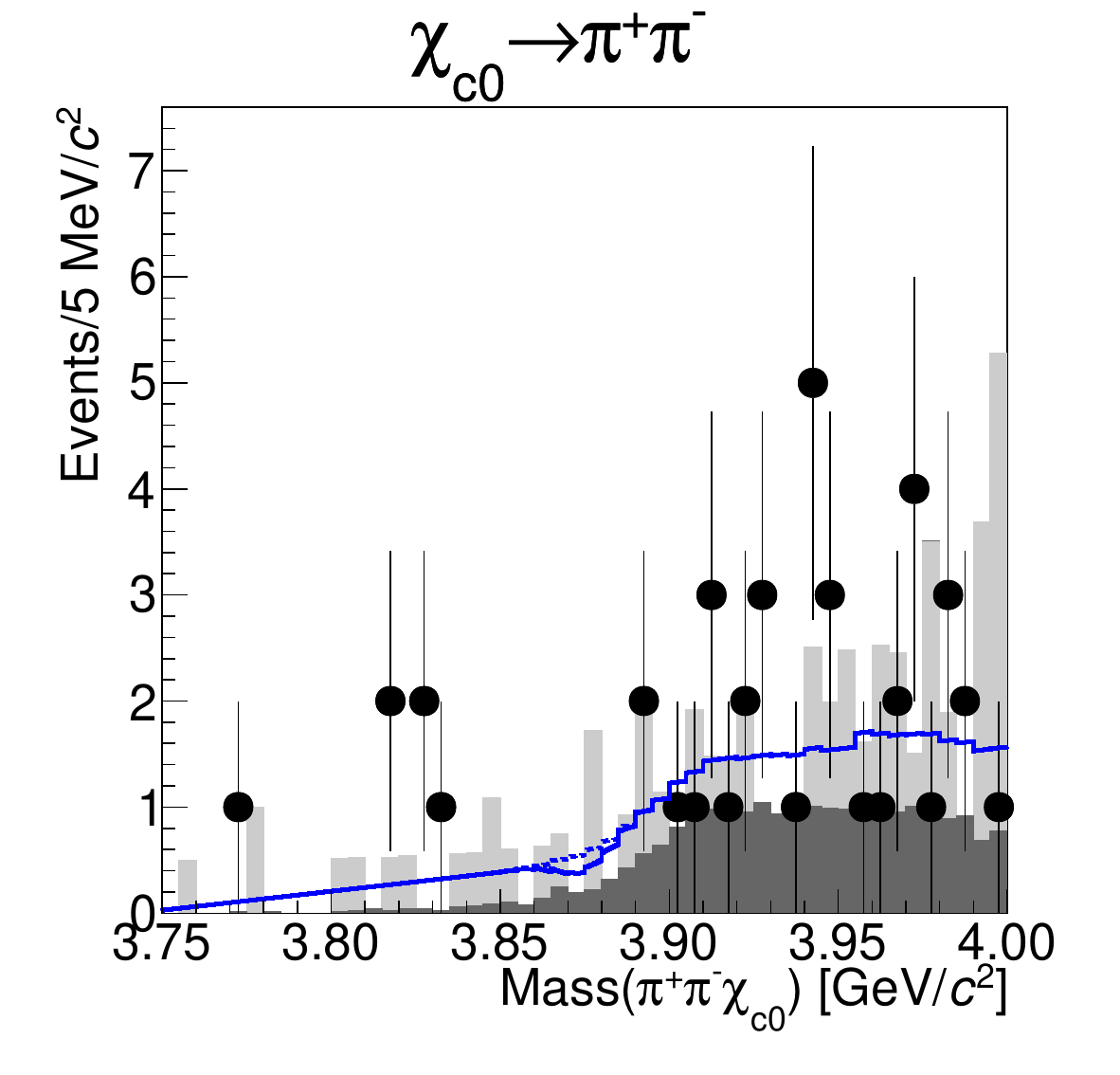}\includegraphics[scale=0.26]{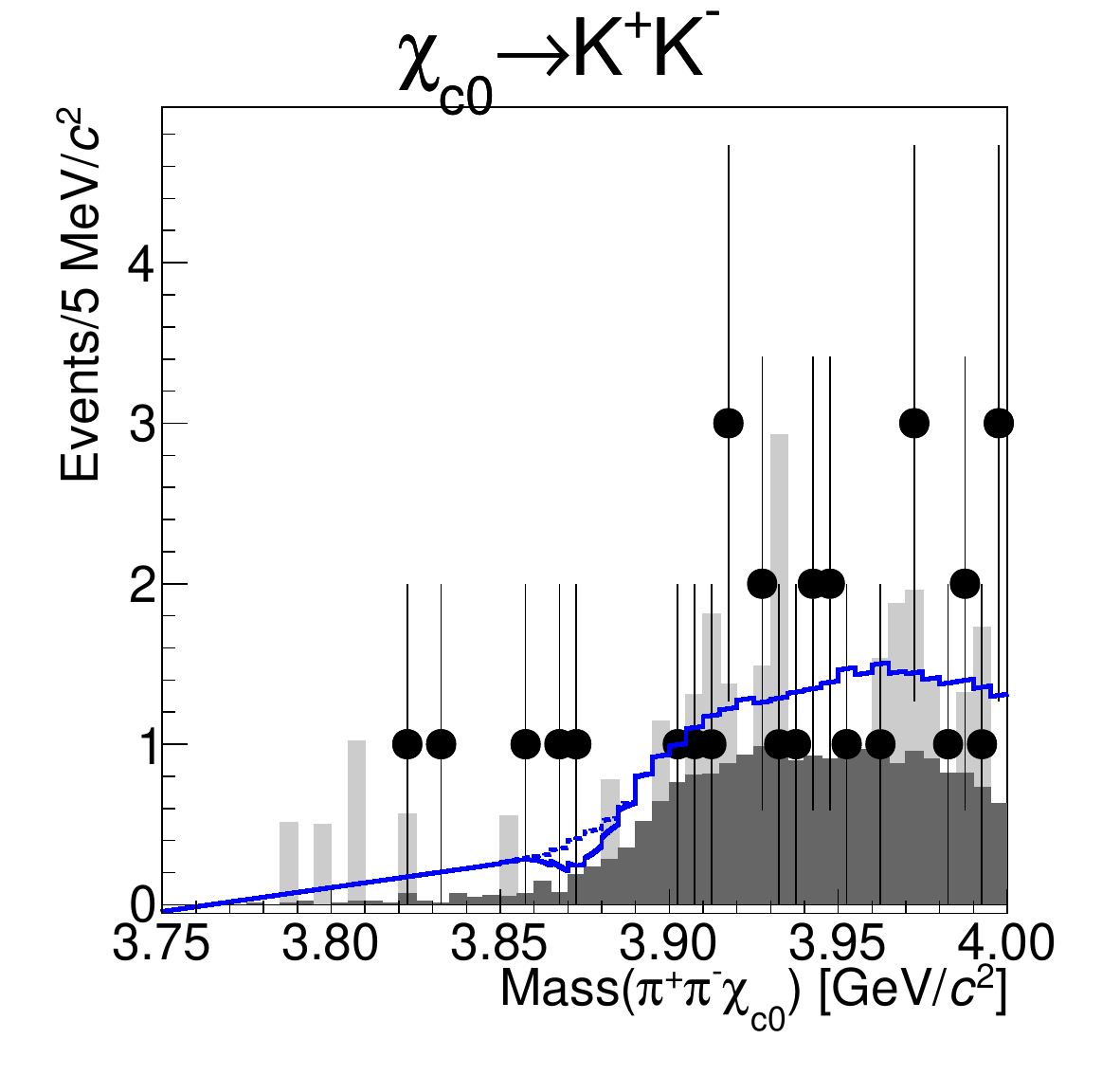}\includegraphics[scale=0.26]{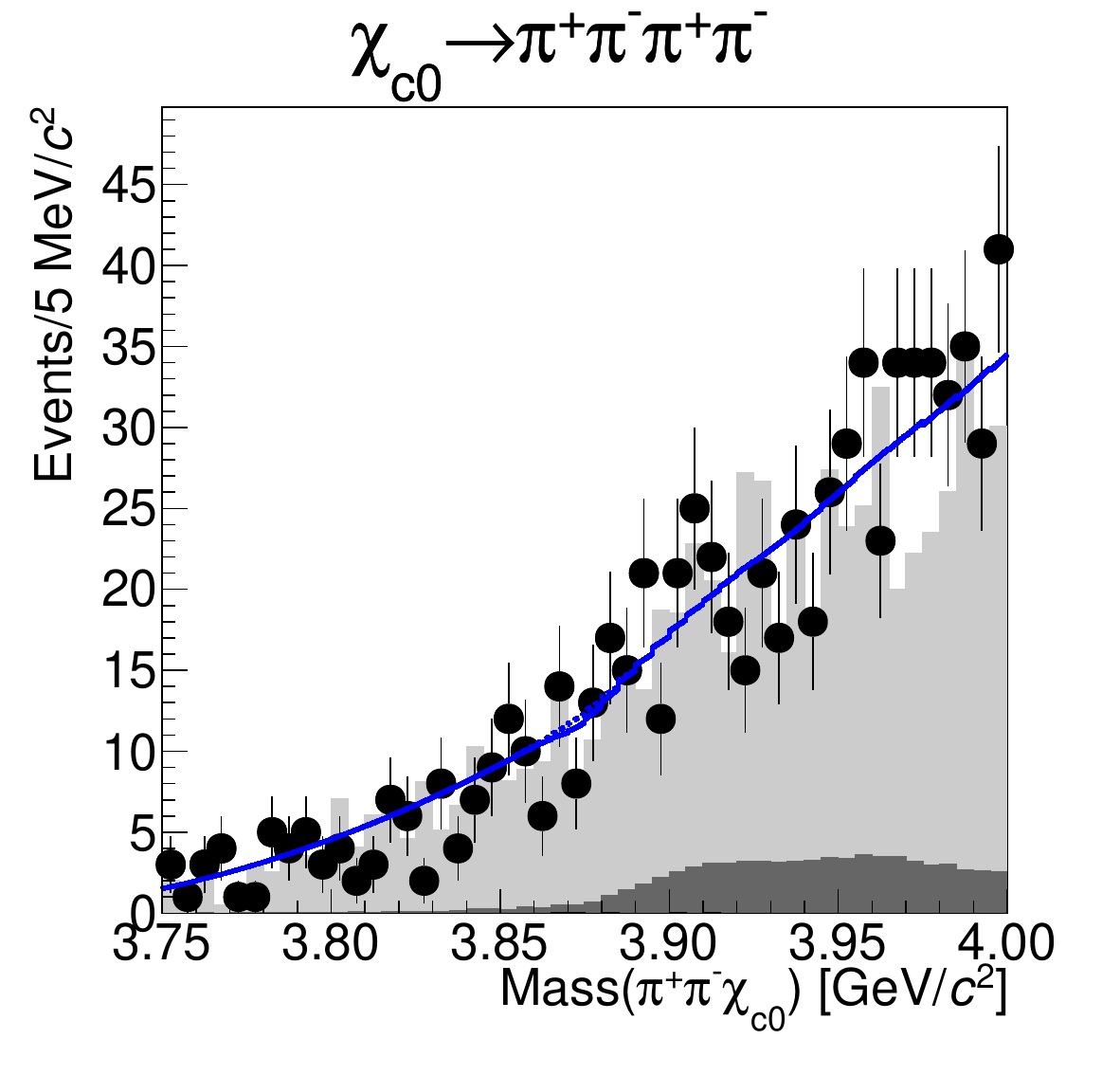}\\\includegraphics[scale=0.26]{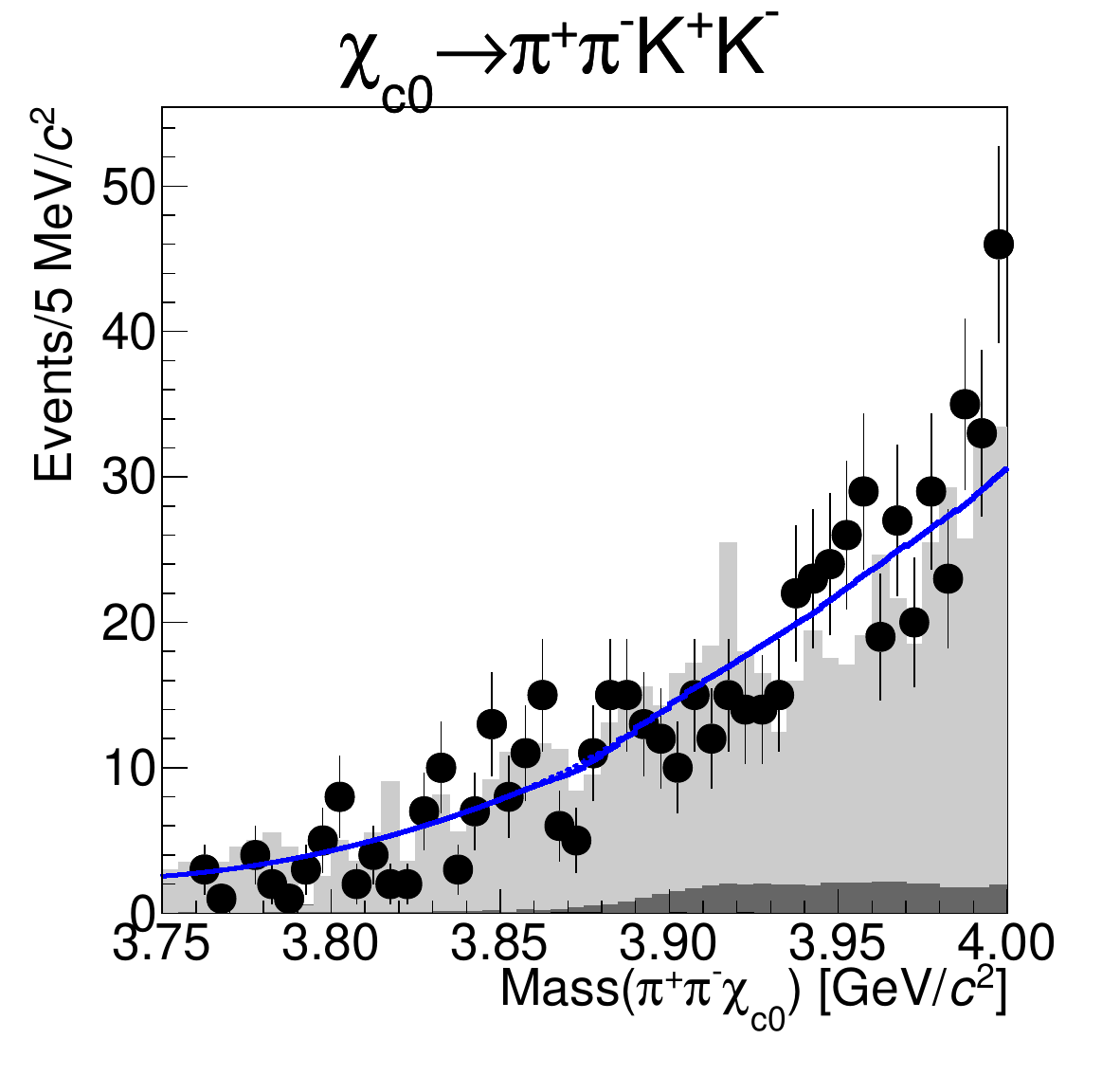}\includegraphics[scale=0.26]{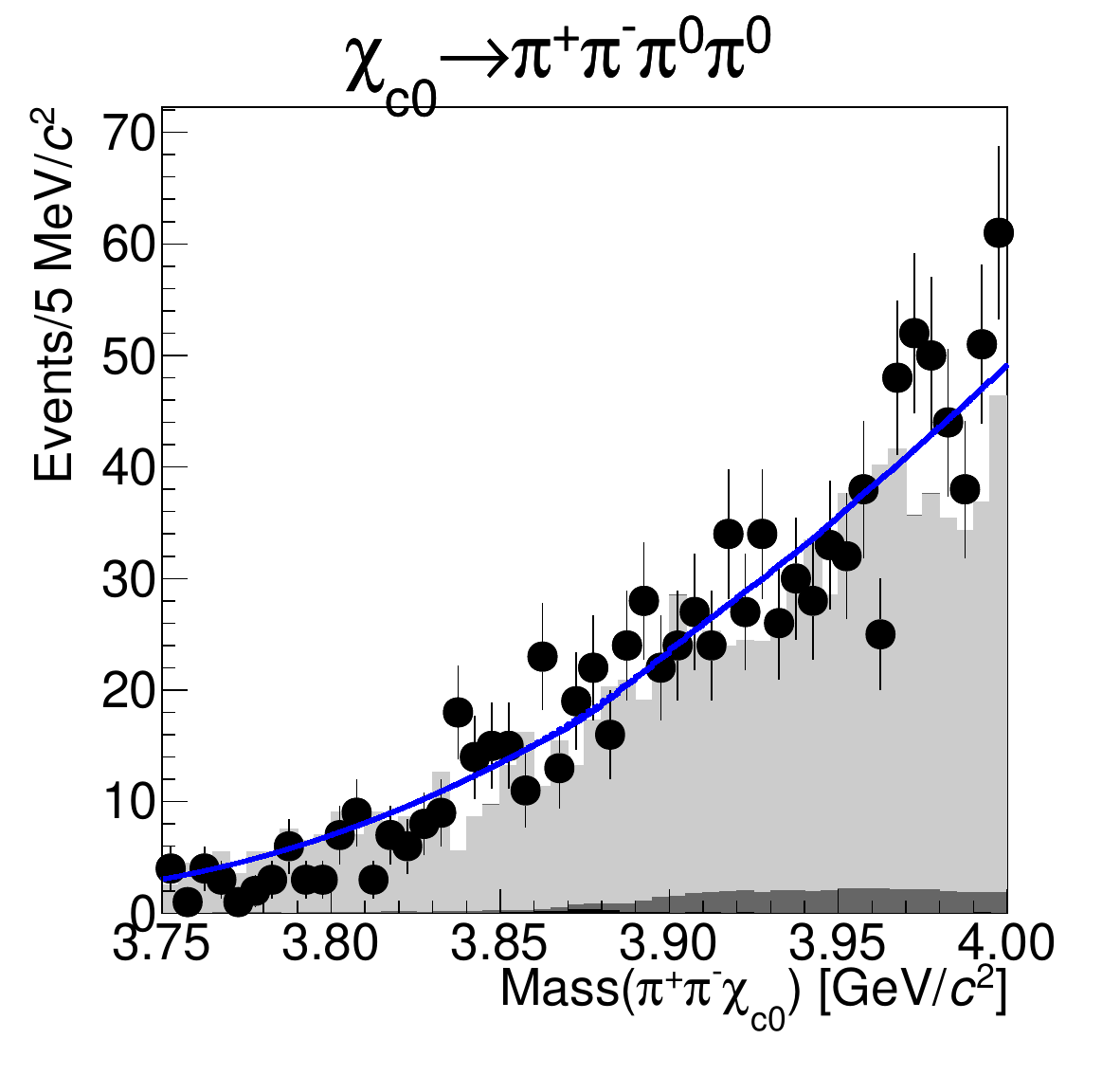}
  \caption{Fit results to the $\pi^+\pi^-\chi_{c0}$ mass spectrum. The points are data in the signal region, the light gray histogram is $\chi_{c0}$ sidebands in data, and the dark gray histogram is the peaking $\chi_{c0}$ background estimated from MC simulated with $e^+e^-\to\pi^+\pi^-\psi(2S)$ with $\psi(2S)\to\gamma\chi_{c0}$. The black histogram is from $X(3872)$ background MC simulations. The solid lines show the fit with a signal component, while the dashed lines are the background functions. There is no evidence of a signal for the $X(3872)$ state.}
  \label{fig:pipi}
\end{figure*}

The fit result for the $\pi^+\pi^-\chi_{c0}$ mass spectrum is shown in Figure~\ref{fig:pipi}. There is no evidence of a signal for this decay mode. The efficiencies, scaling factors, and yields are shown in Table~\ref{tab:pipi}.

\begin{table}
  \caption{ Efficiencies, scaling factors, and signal yields for each decay mode of the $\chi_{c0}$ for $X(3872)\to\pi^+\pi^-\chi_{c0}$. The scaling factors show the relative contribution of each component of the simultaneous fit. The uncertainties on the yields are statistical only. }
  \centering
\begin{tabular}{|c|c|c|c|}
\hline
Decay  &  Efficiency  &  Scale ($w_i$)  &  Signal Yield  \\
\hline
$\chi_{c0}\to\pi^+\pi^-$  &  27.7\%  &  0.00157  &  $-0.59\pm0.88$  \\
\hline
$\chi_{c0}\to K^+ K^-$  &  24.9\%  &  0.00150  &  $-0.57\pm0.85$  \\
\hline
$\chi_{c0}\to \pi^+ \pi^-\pi^+\pi^-$  &  21.0\%  &  0.00492  &  $-1.9\pm2.8$  \\
\hline
$\chi_{c0}\to \pi^+\pi^- K^+ K^-$  &  17.3\%  &  0.00313  &  $-1.2\pm1.8$  \\
\hline
$\chi_{c0}\to \pi^+\pi^- \pi^0\pi^0$  &  8.82\%  &  0.00284  &  $-1.1\pm1.6$  \\
\hline
\end{tabular}
  \label{tab:pipi}
\end{table}

\subsection{$X(3872)\to\pi^0\pi^0\chi_{c0}$}

\begin{figure*}
  \centering
  \includegraphics[scale=0.26]{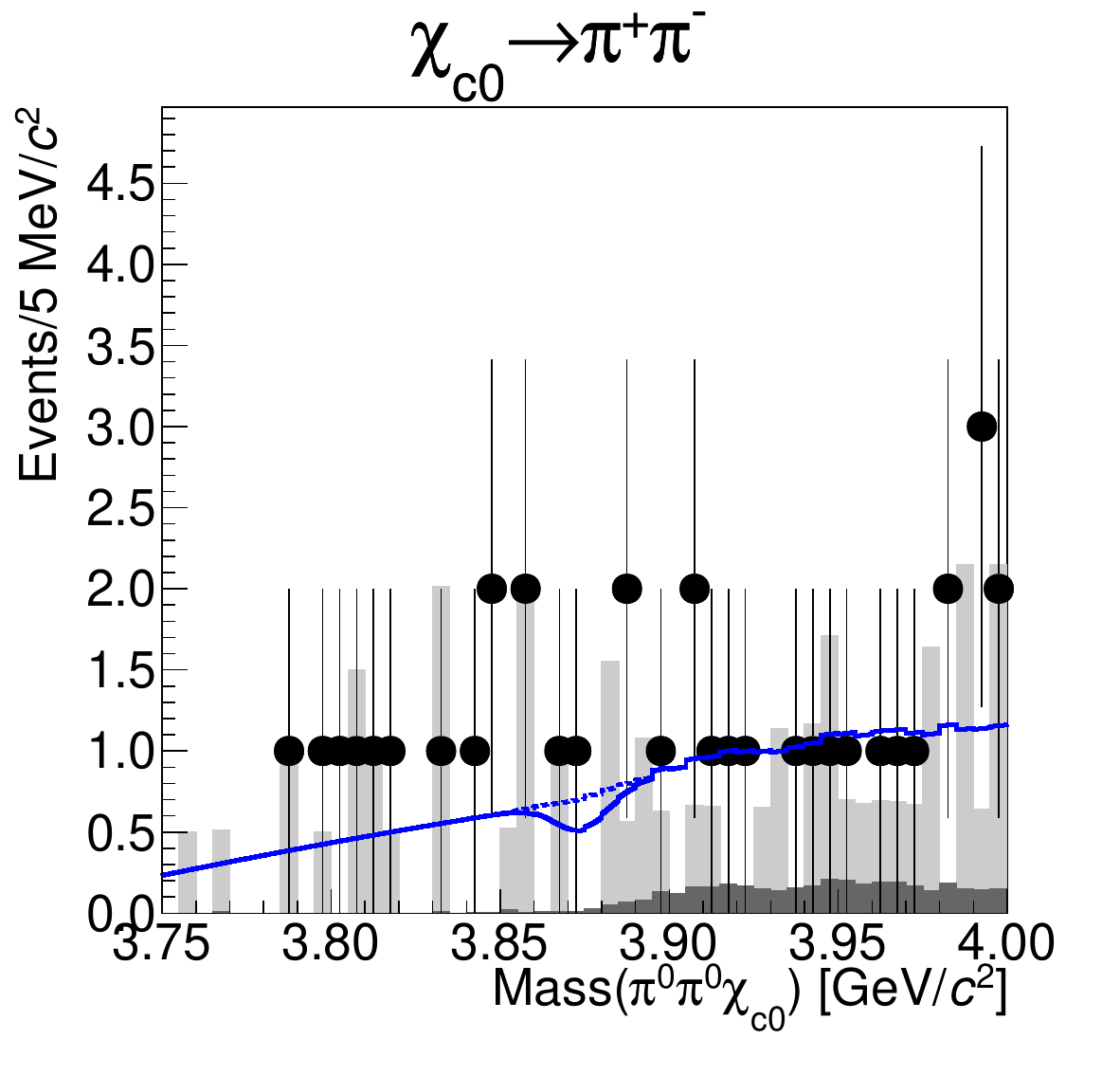}\includegraphics[scale=0.26]{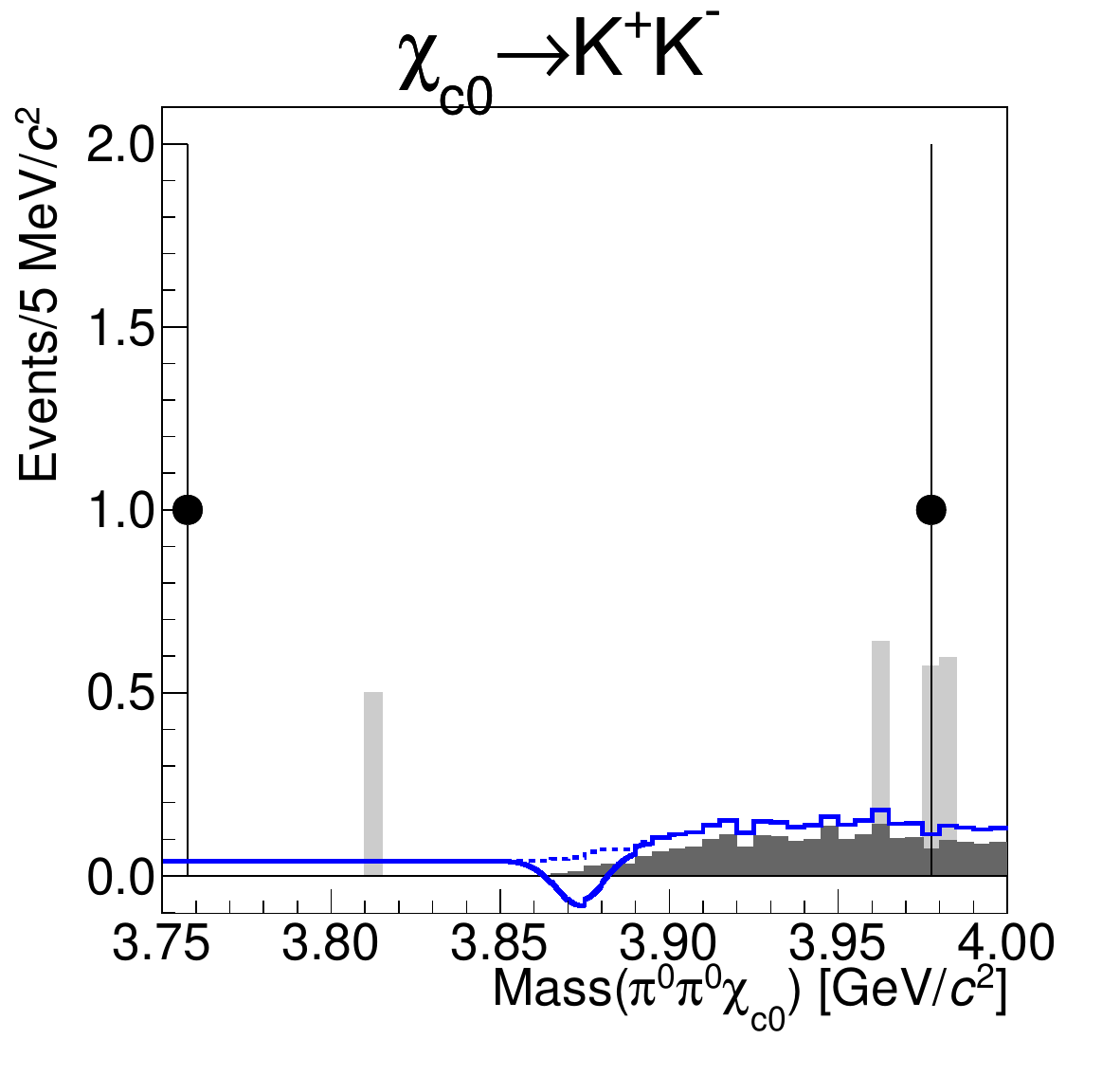}\includegraphics[scale=0.26]{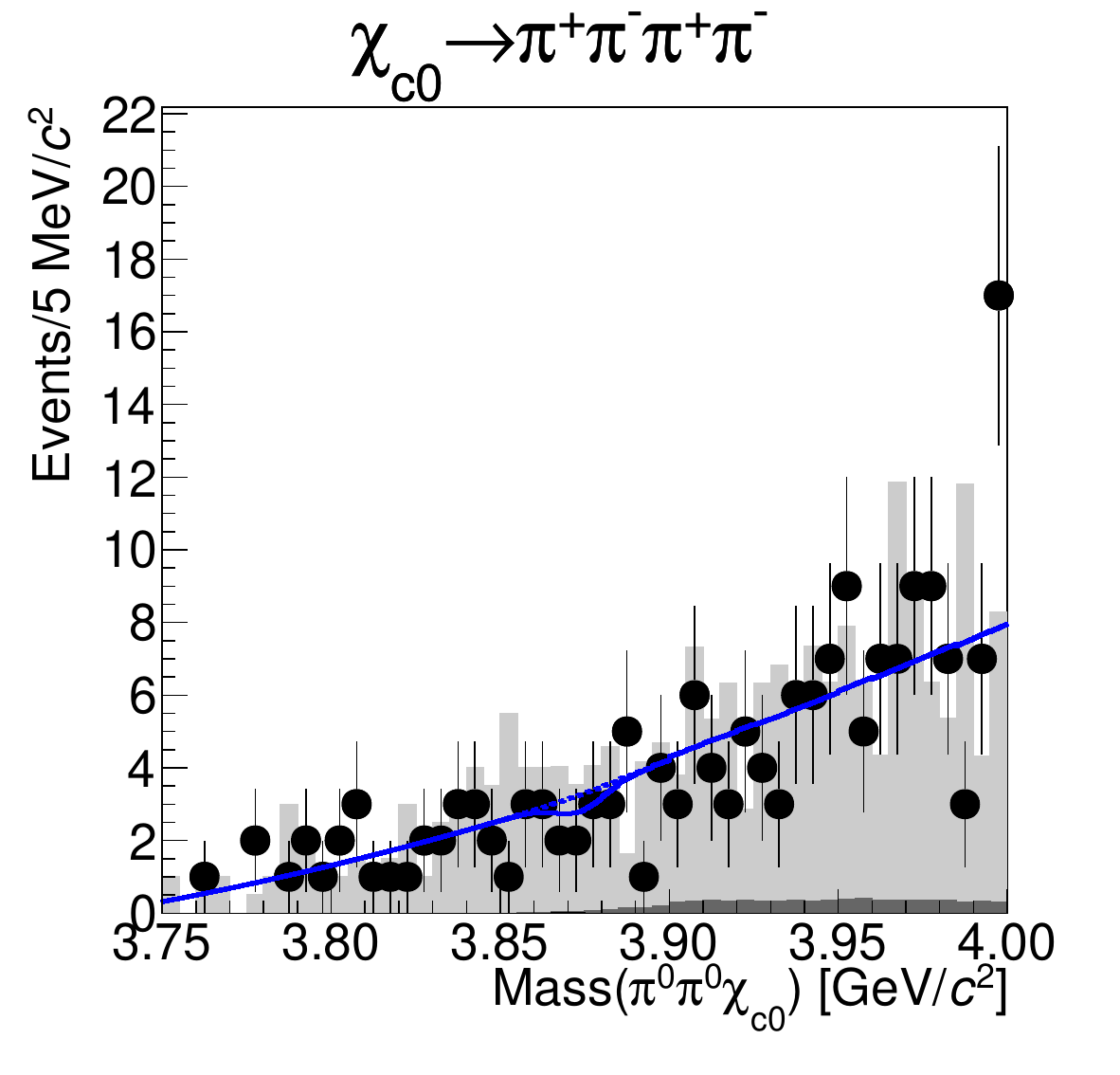}\\\includegraphics[scale=0.26]{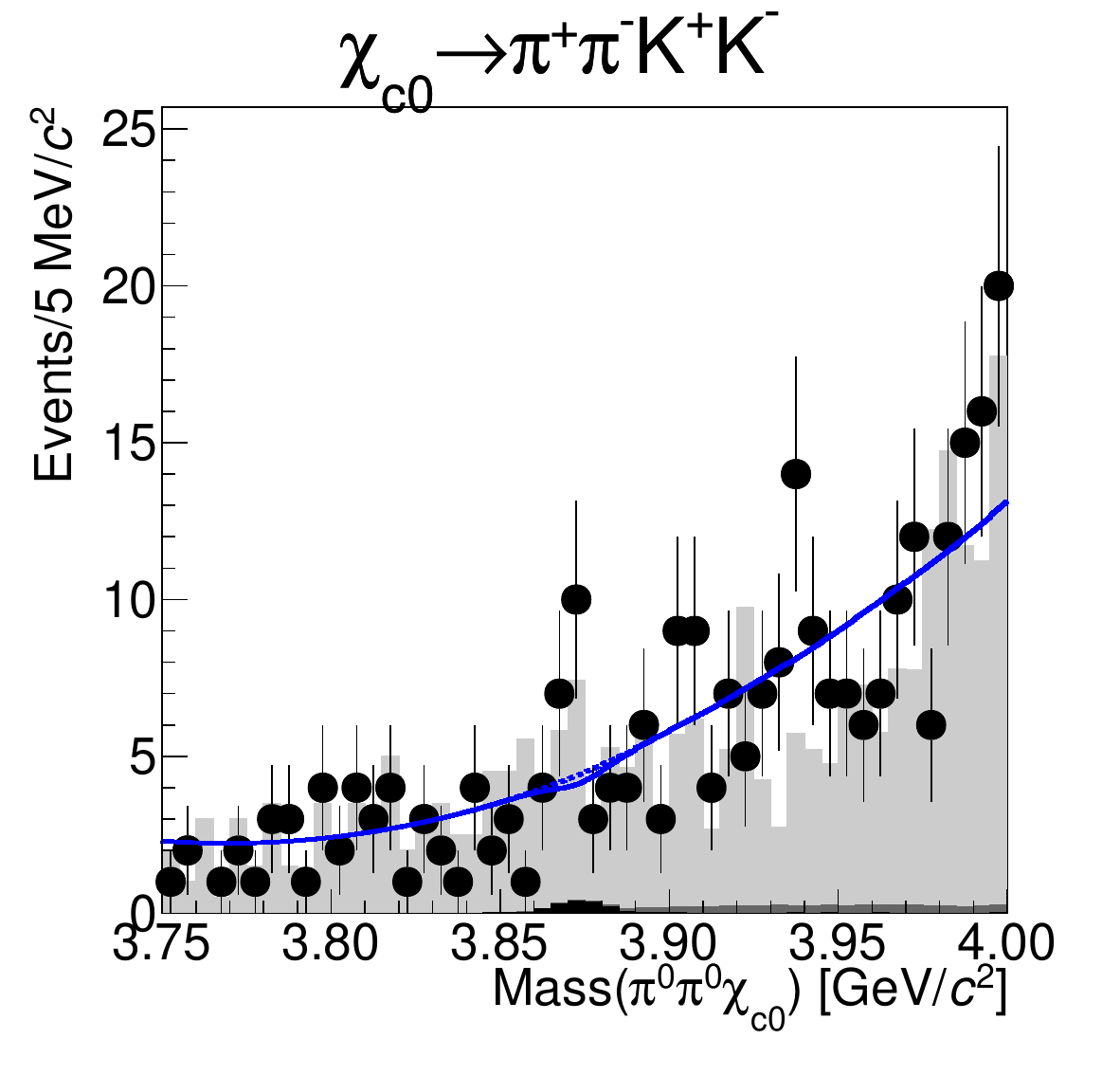}\includegraphics[scale=0.26]{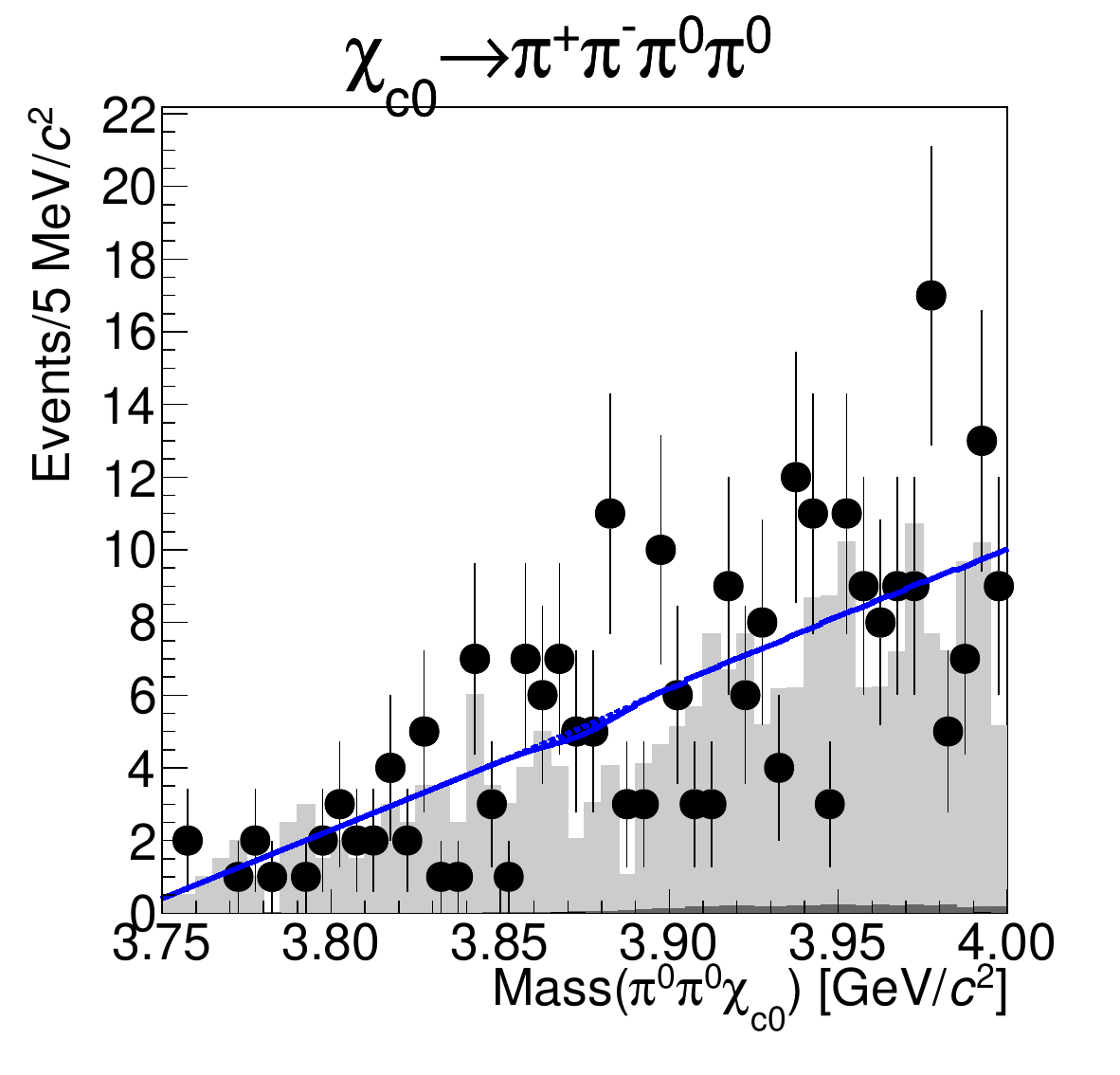}
  \caption{Fit results to the $\pi^0\pi^0\chi_{c0}$ mass spectrum. The points are data in the signal region, the light gray histogram is $\chi_{c0}$ sidebands in data, and the dark gray histogram is the peaking $\chi_{c0}$ background estimated from MC simulated with $e^+e^-\to\pi^0\pi^0\psi(2S)$, with $\psi(2S)\to\gamma\chi_{c0}$. The black histogram is from $X(3872)$ background MC simulations. The solid lines show the fit with a signal component, while the dashed lines are the background functions. There is no evidence of a signal for the $X(3872)$ state.}
  \label{fig:pi0pi0}
\end{figure*}

The fit to the $\pi^0\pi^0\chi_{c0}$ mass spectrum is shown in Figure~\ref{fig:pi0pi0}. There is no evidence for a signal. The efficiencies, scaling factors, and signal yields for $X(3872)\to\pi^0\pi^0\chi_{c0}$ are summarized in Table~\ref{tab:pi0pi0}. Note that $\chi_{c0}\to K^+K^-$ has a much smaller efficiency than $\chi_{c0}\to\pi^+\pi^-$ in this case because of the more stringent $\chi^2/DOF$ requirement shown in Table \ref{tab:optCutpi0}.

\begin{table}
  \caption{ Efficiencies, scaling factors, and signal yields for each decay mode of the $\chi_{c0}$ for $X(3872)\to\pi^0\pi^0\chi_{c0}$. The uncertainties on the yields are statistical only. The scaling factors show the relative contribution of each component of the simultaneous fit.}
  \centering
\begin{tabular}{|c|c|c|c|}
\hline
Decay  &  Efficiency  &  Scale ($w_i$)  &  Signal Yield  \\
\hline
$\chi_{c0}\to\pi^+\pi^-$  &  11.2\%  &  0.000637  &  $-0.8\pm1.3$  \\
\hline
$\chi_{c0}\to K^+ K^-$  &  6.48\%  &  0.000392  &  $-0.49\pm0.79$  \\
\hline
$\chi_{c0}\to \pi^+ \pi^-\pi^+\pi^-$  &  5.66\%  &  0.00132  &  $-1.6\pm2.7$  \\
\hline
$\chi_{c0}\to \pi^+\pi^- K^+ K^-$  &  4.78\%  &  0.000865  &  $-1.1\pm1.7$  \\
\hline
$\chi_{c0}\to \pi^+\pi^- \pi^0\pi^0$  &  2.31\%  &  0.000744  &  $-0.9\pm1.5$  \\
\hline
\end{tabular}
\label{tab:pi0pi0}
\end{table}

\section{Systematic Uncertainties}

We include a 1\% systematic uncertainty for each photon and charged track that does not cancel in the ratio of branching fractions. The remaining systematic uncertainties are discussed below. 

\subsection{Kinematic Fit}

We use the $\chi^2/DOF$ of the kinematic fit to reduce the background for all of the search channels. Since there are no high statistics final states with similar kinematics, we use the procedure in Ref.~\cite{yuping} to determine this systematic uncertainty. Ref~\cite{yuping} found that MC simulation has a significantly narrower $\chi^2$ distribution than in data. Corrections to the track helix parameters of charged particles are used to improve the agreement of the $\chi^2$ distributions between data and MC simulation. We measure the ratio of the search and normalization channel efficiencies before and after these corrections are applied. By taking the ratio, we properly account for the fact that some of the kinematic fit uncertainty cancels in the ratio. We take half the percent difference as the systematic uncertainty due to the kinematic fit.  This results in a systematic uncertainty between 0.3\% and 2.5\% for all the final states.

\subsection{Signal MC Models}

The nominal signal MC simulation includes the decays of the $\chi_{c0}$ into final states both by uniform distributions in phase space and through the most common intermediate states. To test how sensitive the kinematics are to the intermediate states, we generate new signal MC samples in which the $\chi_{c0}$ only decays through the most common intermediate states. The angular distribution of the photon in $e^+e^-\to\gamma X(3872)$ is modified from the nominal E1 transition to a uniform distribution in phase space. We also modify the angular distribution of the decay $X(3872)\to\pi^0\chi_{c0}$ from a P-wave transition to a uniform distribution in phase space. There is no variation for the decays $X(3872)\to\pi\pi\chi_{c0}$, since they are generated with a uniform distribution in phase space in the nominal signal MC simulation, which results in $X(3872)\to\pi^0\chi_{c0}$ having a larger systematic uncertainty. 

These modifications are done for both the search and normalization channels, so we compare the ratio of efficiencies for the nominal MC and for these variations, and take the percent difference as a systematic uncertainty. Normalizing the results to $X(3872)\to\pi^+\pi^-J/\psi$, the systematic uncertainties range from 8.0\% to 11.0\% for $X(3872)\to\pi^0\chi_{c0}$ and from 0.3\% to 3.7\% for $X(3872)\to\pi\pi \chi_{c0}$. When normalizing $X(3872)\to\pi^0\chi_{c0}$ to $X(3872)\to\pi^0\chi_{c1}$, the uncertainty in the simulated model of the denominator must also be taken into account. To account for this, we include the uncertainty of 8.1\% from Ref. \cite{ryan}, which is added in quadrature with the total from $X(3872)\to\pi^0\chi_{c0}$.

\subsection{$\chi_{c0}$ Mass Window}

We select a 50 MeV/$c^2$ window centered on the $\chi_{c0}$ mass when we perform our fits. To test the systematic uncertainty related to this selection, we fit the $\chi_{c0}$ distribution in signal MC simulation using a Voigtian function. For the initial fit, we fix the intrinsic width of the Voigtian to be the intrinsic width of the $\chi_{c0}$ resonance from the PDG~\cite{pdg}, which is 10.8 MeV. As a variation, we widen the intrinsic width of the $\chi_{c0}$ by its PDG uncertainty of 0.6 MeV, and we also widen the width of the Gaussian function by 20\%. A variation of 20\% was chosen because a previous analysis of the process $e^+e^-\to\gamma\eta_c$ at BESIII included a study comparing hadronic final states in MC simulation and data, and it found the resolutions can differ by up to 20\%~\cite{Lara}. We take the percent difference in the number of events from the two fits as the systematic uncertainty. These vary from 1.9\% to 6.0\%.

\subsection{$E_{\rm CM}$ Dependence on the Efficiency Ratio}

The ratio of branching fractions depends on the ratio of efficiencies $\frac{\epsilon(X(3872)\to\pi^0\chi_{c0})}{\epsilon(X(3872)\to\pi^+\pi^-J/\psi)}$. To combine the efficiency ratio measurements at different energies, we perform a weighted average, where the weights are the luminosity times cross section for that energy. The default cross section we use is the $e^+e^-\to\gamma X(3872)$ cross section measured in Ref.~\cite{besX}. To probe the systematic uncertainty due to the $E_{\rm CM}$ dependence of the efficiency ratio, we also use the cross sections for the process $e^+e^-\to\pi^+\pi^-J/\psi$ measured in Ref.~\cite{lineshape}, and one based on the $\psi(4160)$, which is modeled as a Breit-Wigner function with parameters taken from the PDG~\cite{pdg}. We take the largest deviation from the nominal ratio of efficiencies as the systematic uncertainty. This systematic uncertainty varies from 1.6\% to 3.1\%. 

\subsection{Input Branching Fractions}

The branching fractions of the $\chi_{c0}$ decays are used to constrain the relative sizes of the simultaneous fit components. The branching fractions for the decays $\chi_{c0}\to\pi^+\pi^-$, $\chi_{c0}\to K^+K^-$, $\chi_{c0}\to\pi^+\pi^-\pi^+\pi^-$, and $\chi_{c0}\to\pi^+\pi^-K^+K^-$ are included in a constrained fit done by the PDG that included 248 results from previous papers, so the correlated uncertainties between these four branching fractions are known~\cite{pdg}.
To calculate this systematic uncertainty, we generate new sets of input branching fractions using a multivariate Gaussian function, which uses the known values, uncertainties, and correlations as input. For each set of branching fractions, we refit the data with updated fit scales. This procedure is repeated 5000 times for each $X(3872)$ decay, and we measure a new value for the ratio of branching fractions with respect to the decay $X(3872)\to\pi^+\pi^-J/\psi$ for each fit. The resulting distribution of the ratio of branching fractions is then fit with a Gaussian function. The standard deviation of the Gaussian distribution divided by the nominal ratio is the systematic uncertainty, which is \BFUNCONE~for $X(3872)\to\pi^0\chi_{c0}$, \BFUNCTWO~for $X(3872)\to\pi^+\pi^-\chi_{c0}$, and \BFUNCTHREE~for $X(3872)\to\pi^0\pi^0\chi_{c0}$ decays.

\subsection{Fit Model}

The last systematic uncertainty is due to the fit model. We vary seven parameters: 1) the non-$\chi_{c0}$ background is parameterized with a first, second or third order polynomial; 2) the central value for the $X(3872)$ is taken from signal Monte Carlo or varied $\pm 1$ MeV \cite{pdg}; 3) the internal width is varied from 0.96 MeV \cite{narrow1} to 1.39 MeV \cite{narrow2}; 4) the resolution of the Gaussian core of the Voigtian is taken from Monte Carlo or increased by 20\% \cite{Lara}; 5) the fit range is narrowed from [3.75, 4.0] GeV$/c^2$ to [3.775, 3.975] GeV/$c^2$; 6) the beam energy is taken as the nominal measurement or varied $\pm$ 1 MeV \cite{ecm1,ecm2}; 7) the size of the background Monte Carlo is varied by its uncertainty $\sigma_{mc}$, which is determined varying the resonance parameters in the fits to $\sigma(e^+e^-\to\omega\chi_{c0})$ \cite{omegaChi} or $\sigma(e^+e^-\to\pi\pi\psi(2S))$ \cite{pipiPsi2S} and determining the relative uncertainty on the number of expected events produced (5.8\% for $\omega\chi_{c0}$, 20\% for $\pi\pi\psi(2S)$). This results in a total of $3\cdot3\cdot2\cdot2\cdot2\cdot3\cdot3=648$ fit variations. The largest upper limit out of all of these variations is reported. The list of fit variations is summarized in Table \ref{tab:variations}.

\begin{table}[ht]
  \caption{List of all the fit variations used in this analysis. Variations to each of these components are done independently, so the total number of variations is $3\cdot3\cdot2\cdot2\cdot2\cdot3\cdot3=648$.}
  \centering
  \begin{tabular}{|c|c|c|c|}\hline
   & Variation & Description & Variations\\\hline
   1 & Polynomial order &  1st, 2nd, 3rd order & 3\\\hline
   2 &  Mass & Nominal, and $\pm 1$ MeV & 3 \\\hline
   3 & Internal width & 0.96 MeV, 1.39 MeV & 2\\\hline
   4 & Resolution &  Nominal, widen 20\% & 2\\\hline
   5 & Fit range & Nominal, $3.775-3.975$ GeV & 2 \\\hline
   6 & Beam energy & Nominal, $\pm 1$ MeV & 3\\\hline
   7 & MC Scale & Nominal, $\pm\sigma_{mc}$ & 3\\\hline
 \end{tabular}

   \label{tab:variations}
 \end{table}

\subsection{Total Systematic Uncertainties}

Several of the systematic uncertainties are correlated between the different $\chi_{c0}$ decay modes. For these uncertainties, we calculate a weighted average to get the total systematic uncertainty from that source, using the efficiency times branching fraction as the weights. This is done for the tracking, photon, kinematic fit, MC simulation model, $\chi_{c0}$ selection, and $\epsilon$ ratio systematic uncertainties. We summarize the total systematic uncertainty values for $X(3872)\to\pi^0\chi_{c0}$ in Table~\ref{tab:search}, and $X(3872)\to\pi^+\pi^-\chi_{c0}$ and $X(3872)\to\pi^0\pi^0\chi_{c0}$ in Table~\ref{tab:pipiSys}. As previously mentioned, the fitting uncertainty is included for the upper limits by using the variation that results in the largest upper limit.

\begin{table}
    \caption{Total systematic uncertainties for the decay $X(3872)\to\pi^0\chi_{c0}$. Here we show the systematic uncertainty when the branching fraction is normalized to $X(3872)\to\pi^+\pi^-J/\psi$ as well as $X(3872)\to\pi^0\chi_{c1}$. This is done because some of the photon and charged track systematic uncertainties cancel in the ratio of branching fractions. }
  \begin{tabular}{|c|c|c|}
\hline
Source  &  $R_{\pi^+\pi^-J/\psi}$ Total  &  $R_{\pi^0\chi_{c1}}$ Total  \\
\hline
Tracking  &  5.1\%  &  3.7\%  \\
\hline
Photon Efficiency  &  2.8\%  &  1.3\%  \\
\hline
$\chi^2/DOF$ Cut  &  1.2\%  &  1.2\%  \\
\hline
Decay Models  &  9.5\%  &  12.5\%  \\
\hline
Branching Fractions  &  4.7\%  &  4.7\%  \\
\hline
$\chi_{c0}$ Selection  &  3.3\%  &  3.3\%  \\
\hline
$\epsilon$ Ratio  &  2.6\%  &  2.6\%  \\
\hline
Total  &  12.8\%  &  14.5\%  \\
\hline
\end{tabular}
\label{tab:search}
\end{table}

\begin{table}
    \caption{Total systematic uncertainties for $X(3872)\to\pi^+\pi^-\chi_{c0}$ and $X(3872)\to\pi^0\pi^0\chi_{c0}$ modes.}
  \begin{tabular}{|c|c|c|}
\hline
Source  &  $X(3872)\to\pi^+\pi^-\chi_{c0}$ & $X(3872)\to\pi^0\pi^0\chi_{c0}$  \\
\hline
Tracking  &  3.7\% & 5.1\% \\
\hline
Photons  &  0.8\% & 4.8\% \\
\hline
$\chi^2/DOF$ Cut  &  0.7\% & 1.1\%  \\
\hline
Decay Models  &  2.3\% & 1.9\% \\
\hline
$\mathcal{B}(\chi_{c0})$ &  \BFUNCTWO & \BFUNCTHREE \\
\hline
$\chi_{c0}$ Selection  &  2.9\% & 3.3\%  \\
\hline
$\epsilon$ Ratio  &  2.3\% & 2.0\% \\
\hline
Total  &  \PIPISYSUNC & \TWOPIZEROSYSUNC \\
\hline
\end{tabular}
\label{tab:pipiSys}
\end{table}

\section{\label{sec:level1} Calculation of Upper Limits}

\begin{figure*}
  \centering
  \includegraphics[scale=0.3]{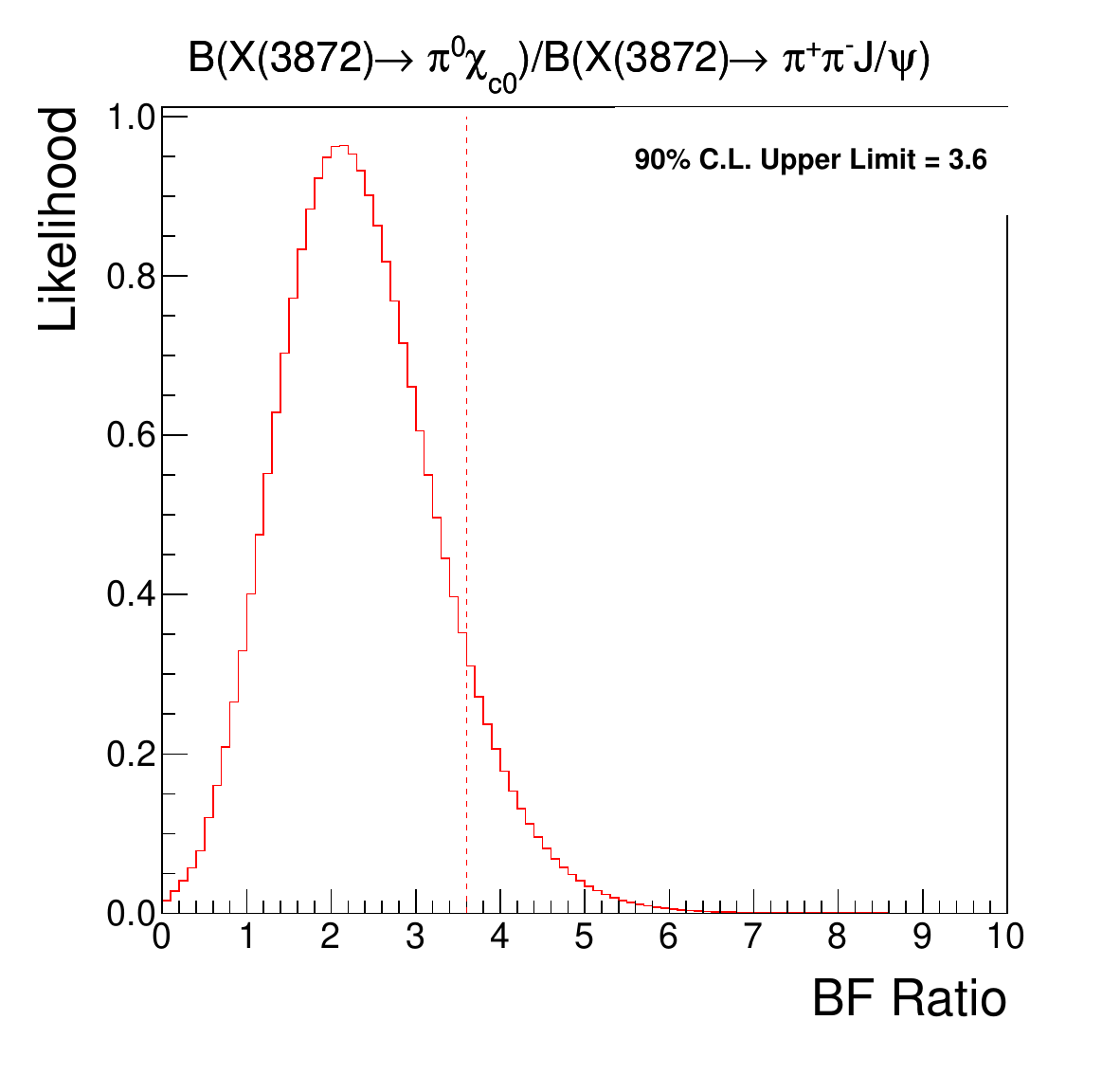}\includegraphics[scale=0.3]{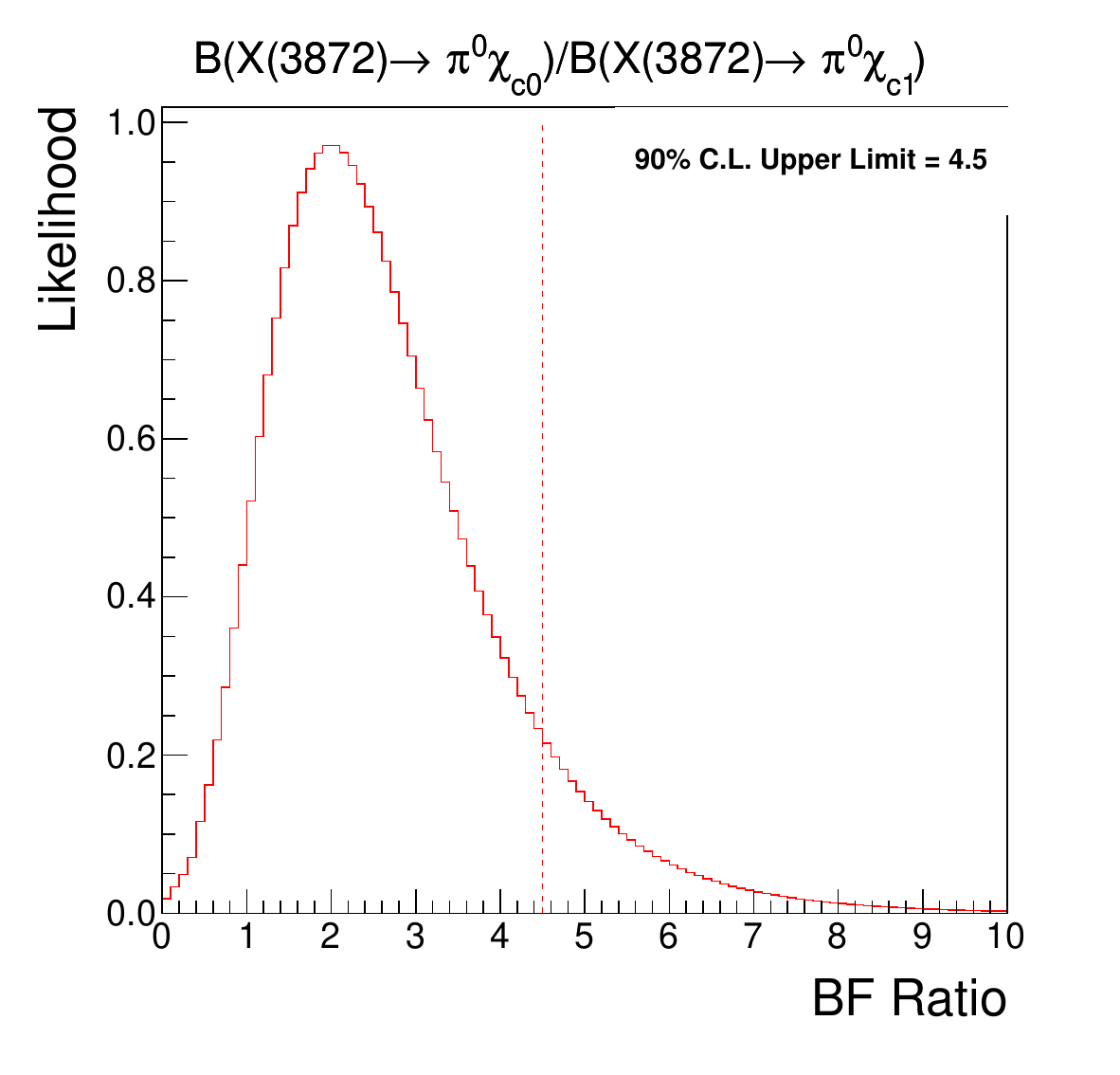}\\
  \includegraphics[scale=0.3]{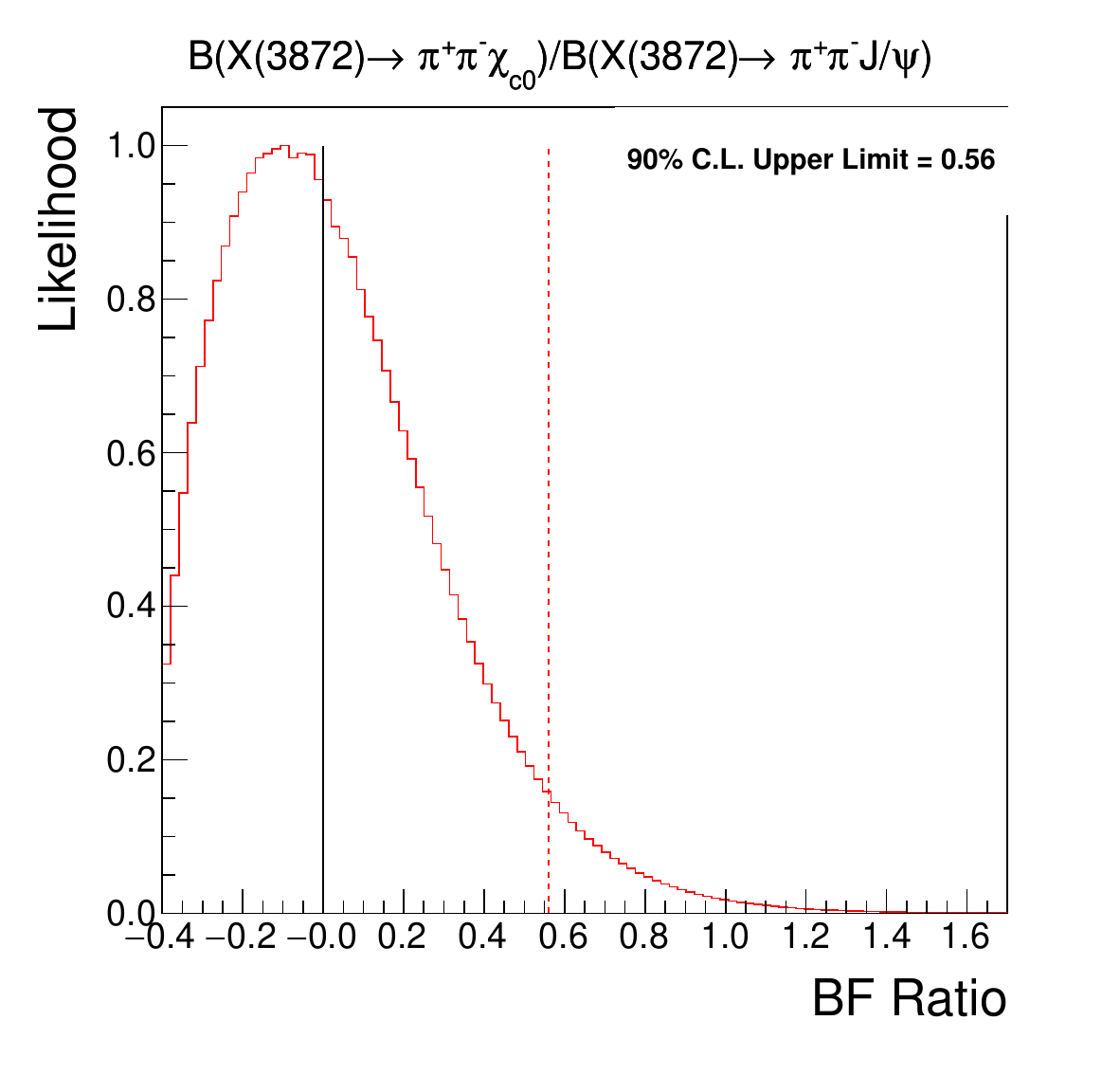}\includegraphics[scale=0.3]{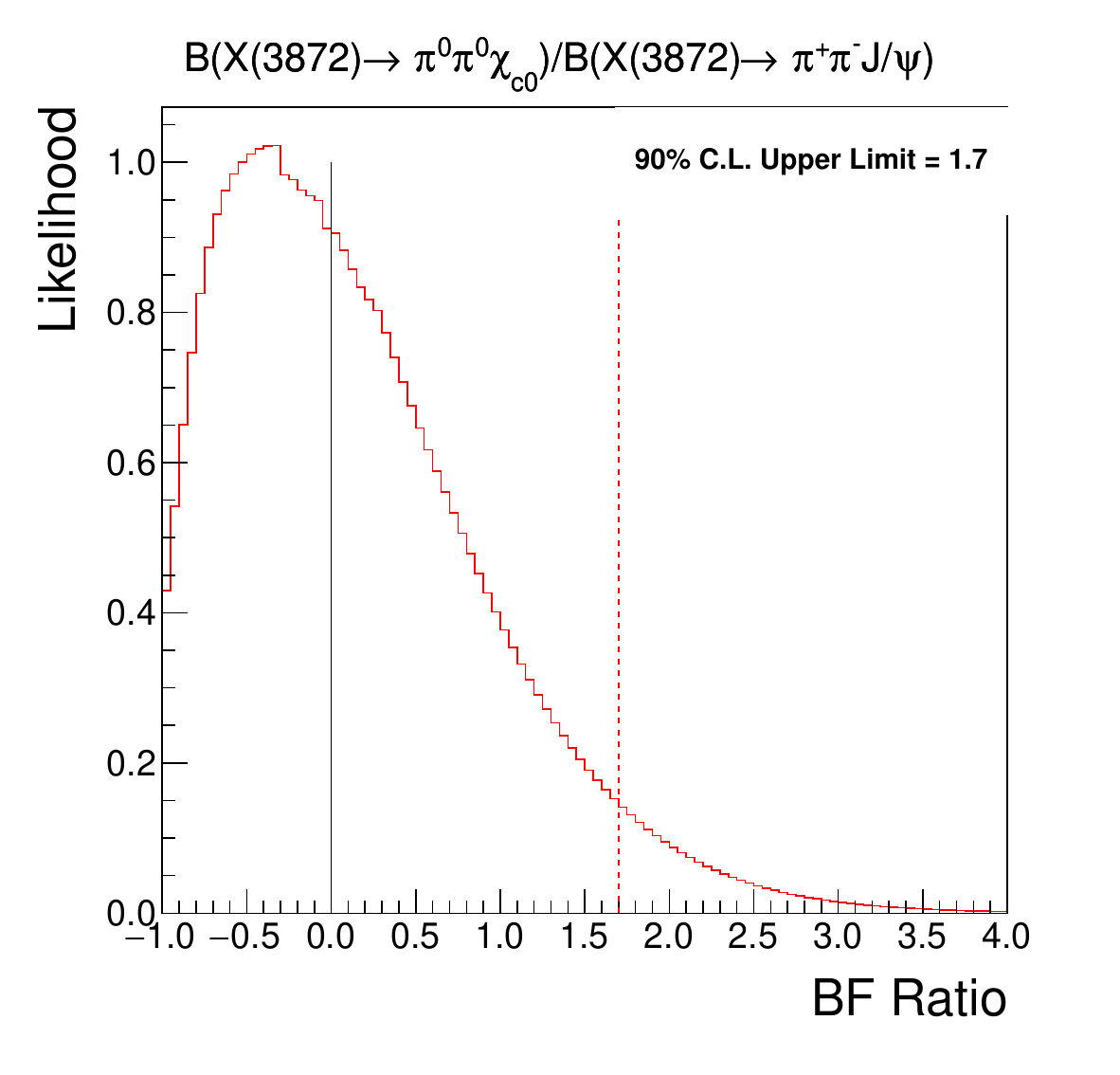}
  \caption{Plots showing the likelihood scans for all four measured ratios of branching fractions. The top left shows \RMAIN, top right shows \RDOUBLE, bottom left shows \RTWOPI$\textrm{ }$and bottom right shows \RTWOPIZERO. The histogram is the likelihood curve that gives the largest upper limit after convolving by the systematic uncertainty. The dashed line shows the 90\% C.L. upper limit value. }
\label{fig:likelihoods}
\end{figure*}

Our upper limits have to include the statistical uncertainty from both the numerator and denominator of the ratio \RMAIN. To get the total statistical uncertainty correct, we perform a likelihood scan for the search and normalization channels. This is done by performing several hundred fits with the yield fixed to different values and with the background parameters floating. Once we have both likelihood distributions, we randomly sample them both to determine new yield values. These yield values are then used to calculate a new value for the ratio of branching fractions. This process is done a million times, and the resulting distribution gives us the likelihood for the ratio of branching fractions. To determine the upper limit, we use the likelihood function that results in the largest upper limit. We convolve this likelihood function with a Gaussian function that has a width corresponding to the systematic uncertainty. The upper limits are calculated by integrating the resulting curve from 0 up to the point where 90\% of the distribution is below the upper limit. Figure~\ref{fig:likelihoods} shows the likelihood curves for the largest upper limit (red histogram) after including the systematic uncertainties, as well as the upper limit values (vertical line). 

The fit variation that results in the largest upper limit for $X(3872)\to\pi^0\chi_{c0}$ is a fit with a first order polynomial function where the signal function has a mass shifted $+1$ MeV$/c^2$ from the nominal value, an internal width of 1.39 MeV, a resolution increased by 20\%, a beam energy shifted $-1$ MeV, and the $\omega\chi_{c0}$ background scale decreased by 5.8\%. For $X(3872)\to\pi^+\pi^-\chi_{c0}$, the variation that results in the largest upper limit is a fit with a second order polynomial function where the signal function has an increased resolution by 20\%, fit range narrowed by 50 MeV/$c^2$, a beam energy shifted $-1$ MeV, and the $\pi\pi\psi(2S)$ background scale increased by 20\%. For $X(3872)\to\pi^0\pi^0\chi_{c0}$, the variation that results in the largest upper limit is a fit with a third order polynomial function where the signal function has a mass shifted $-1$ MeV/$c^2$, an increased resolution by 20\%, and the $\pi\pi\psi(2S)$ background scale increased by 20\%.

\section{Summary}

In summary, we perform a search for the decays $X(3872)\to\pi^0\chi_{c0}$, $X(3872)\to\pi^+\pi^-\chi_{c0}$, and $X(3872)\to\pi^0\pi^0\chi_{c0}$. The significance for $X(3872)\to\pi^0\chi_{c0}$ is found to be at least \SIGNIFMIN$\textrm{ }$. Since the fit results are all below 3$\sigma$, we set an upper limit of \RMAIN$<$ \UL~at 90\% C.L, which is a significant improvement over the previous BESIII upper limit of \RMAIN$<19$ \cite{ryan}. Combined with the previous BESIII result~\cite{ryan} of \RRYAN$=0.88^{+0.33}_{-0.27}\pm0.10$, we set an upper limit of \RDOUBLE$<$ \FINALUL~at 90\% C.L. This upper limit is too large to rule out any interpretation of the $X(3872)$ state. 

We find no significant signals for $X(3872)\to\pi^+\pi^-\chi_{c0}$ and $X(3872)\to\pi^0\pi^0\chi_{c0}$, so we set upper limits of \RTWOPI$<\PIPIUL$, and \RTWOPIZERO$<\TWOPIZEROUL $ at 90\% C.L. Both of them are consistent with theoretical predictions from Ref.~\cite{voloshin} that they should be suppressed regardless of whether the $X(3872)$ is a four-quark or charmonium state. All of our results are summarized in Table~\ref{tab:final}.

\begin{table}[ht]
  \caption{90\% C.L. upper limits for each of our measurements.}
  \centering
\begin{tabular}{ |c|c| }
\hline
Ratio & 90\% C.L Upper Limit \\ \hline
{\Large \RMAIN }& \UL\\ \hline
 {\Large \RDOUBLE} & \FINALUL \\ \hline
 {\Large \RTWOPI }& \PIPIUL \\ \hline
  {\Large \RTWOPIZERO} & \TWOPIZEROUL \\\hline
\end{tabular}
\label{tab:final}
\end{table}

\section{Acknowledgments}

The BESIII collaboration thanks the staff of BEPCII and the IHEP computing center for their strong support. This work is supported in part by National Key R\&D Program of China under Contracts Nos. 2020YFA0406300, 2020YFA0406400; National Natural Science Foundation of China (NSFC) under Contracts Nos. 11625523, 11635010, 11735014, 11822506, 11835012, 11935015, 11935016, 11935018, 11961141012, 12022510, 12025502, 12035009, 12035013, 12061131003; the Chinese Academy of Sciences (CAS) Large-Scale Scientific Facility Program; Joint Large-Scale Scientific Facility Funds of the NSFC and CAS under Contracts Nos. U1732263, U1832207; CAS Key Research Program of Frontier Sciences under Contract No. QYZDJ-SSW-SLH040; 100 Talents Program of CAS; INPAC and Shanghai Key Laboratory for Particle Physics and Cosmology; ERC under Contract No. 758462; European Union Horizon 2020 research and innovation programme under Contract No. Marie Sklodowska-Curie grant agreement No 894790; German Research Foundation DFG under Contracts Nos. 443159800, Collaborative Research Center CRC 1044, FOR 2359, FOR 2359, GRK 214; Istituto Nazionale di Fisica Nucleare, Italy; Ministry of Development of Turkey under Contract No. DPT2006K-120470; National Science and Technology fund; Olle Engkvist Foundation under Contract No. 200-0605; STFC (United Kingdom); The Knut and Alice Wallenberg Foundation (Sweden) under Contract No. 2016.0157; The Royal Society, UK under Contracts Nos. DH140054, DH160214; The Swedish Research Council; U. S. Department of Energy under Contracts Nos. DE-FG02-05ER41374, DE-SC-0012069

\pagebreak


\clearpage
\begin{thebibliography}{9}
\bibitem{firstObs}
  S.-K. Choi \textit{et al.} (Belle Collaboration), \href{https://doi.org/10.1103/PhysRevLett.91.262001}{Phys. Rev. Lett. \textbf{91}, 262001 (2003).} 

\bibitem{numbers}
    R. Aaij \textit{et al.} (LHCb Collaboration), \href{https://doi.org/10.1103/PhysRevLett.110.222001}{Phys. Rev. Lett. \textbf{110}, 222001 (2013).}
  
\bibitem{chic}
  S. Godfrey and N. Isgur, \href{https://doi.org/10.1103/PhysRevD.32.189}{Phys Rev. D \textbf{32} 189 (1985).}
  

 \bibitem{narrow1}
   R. Aaij \textit{et al.} (LHCb Collaboration), \href{https://doi.org/10.1007/JHEP08(2020)123}{JHEP \textbf{08} 123 (2020).}
   
 \bibitem{narrow2}
   R. Aaij \textit{et al.} (LHCb Collaboration), \href{https://doi.org/10.1103/PhysRevD.102.092005}{Phys. Rev. D \textbf{102}, 092005 (2020).}

  \bibitem{rho}
    S.-K. Choi \textit{et al.} (Belle Collaboration) \href{https://doi.org/10.1103/PhysRevD.84.052004}{Phys. Rev. D \textbf{84}, 052004 (2011).}

  \bibitem{dd}
    T. Aushev \textit{et al.} (Belle Collaboration) \href{https://doi.org/10.1103/PhysRevD.81.031103}{Phys. Rev. D \textbf{81}, 031103 (2010)}; B. Aubert \textit{et al.} (BABAR Collaboration), \href{https://doi.org/10.1103/PhysRevD.77.011102}{Phys. Rev. D \textbf{77}, 011102 (2008).}
    
  \bibitem{gamma}
    R. Aaij \textit{et al.} (LHCb Collaboration), \href{https://doi.org/10.1016/j.nuclphysb.2014.06.011}{Nucl. Phys. \textbf{B886}, 665 (2014)}; V. Bhardwaj \textit{et al.} (Belle Collaboration), \href{https://doi.org/10.1103/PhysRevLett.107.091803}{Phys. Rev. Lett. \textbf{107}, 091803 (2011)}; B. Aubert \textit{et al.} (BABAR Collaboration) \href{https://doi.org/10.1103/PhysRevLett.102.132001}{Phys. Rev. Lett. \textbf{102}, 132001 (2009).}

    \bibitem{ryan}
  M. Ablikim \textit{et al.} (BESIII Collaboration), \href{https://doi.org/10.1103/PhysRevLett.122.202001}{Phys. Rev. Lett. \textbf{122}, 202001 (2019).}

  \bibitem{besX}
    M. Ablikim \textit{et al.} (BESIII Collaboration), \href{https://doi.org/10.1103/PhysRevLett.122.232002}{Phys. Rev. Lett. \textbf{122} 232002 (2019).}

    

  \bibitem{pdg}
    P. A. Zyla \textit{et al.} (Particle Data Group), \href{https://doi.org/10.1093/ptep/ptaa104}{Prog. Theor. Exp. Phys. \textbf{2020}, 083C01 (2020) and 2021 update.}
    
\bibitem{voloshin}
  S. Dubynskiy and M. B. Voloshin,  \href{https://doi.org/10.1103/PhysRevD.77.014013}{Phys. Rev. D. \textbf{77}, 014013 (2008).}


\bibitem{mehen}
  S. Fleming and T. Mehen, \href{https://doi.org/10.1103/PhysRevD.78.094019}{Phys. Rev. D. \textbf{78}, 094019 (2008). }

\bibitem{wu}
  Q. Wu, D. Y. Chen, and T. Matsuki, \href{https://doi.org/10.1140/epjc/s10052-021-08984-2}{Eur. Phys. J. C \textbf{81}, 193 (2021).}


  \bibitem{dong}
  Y. Dong, A. Faessler, T. Gutsche, S. Kovalenko, and V. E. Lyubovitskij, \href{https://doi.org/10.1103/PhysRevD.79.094013}{Phys Rev D \textbf{79} 094013 (2009).}
  
\bibitem{zhou}
  Z. Y. Zhou, M. T. Yu, and Z. Xiao, \href{https://doi.org/10.1103/PhysRevD.100.094025}{Phys. Rev. D. \textbf{100}, 094025 (2019).}


  
\bibitem{belleSearch}
  V. Bhardwaj \textit{et al.} (Belle Collaboration), \href{https://doi.org/10.1103/PhysRevD.99.111101}{Phys. Rev. D, \textbf{99} 111101(R) (2019).}


\bibitem{ecm1}
  M. Ablikim \textit{et al.} (BESIII Collaboration), \href{https://doi.org/10.1088/1674-1137/40/6/063001}{Chinese Phys. C. \textbf{40} 063001 (2016).}

\bibitem{ecm2}
  M. Ablikim \textit{et al.} (BESIII Collaboration), \href{https://doi.org/10.1088/1674-1137/ac1575}{Chinese Phys. C. \textbf{45} 103001 (2021).}

  \bibitem{lum1}
  M. Ablikim \textit{et al.} (BESIII Collaboration), \href{https://doi.org/10.1088/1674-1137/39/9/093001}{Chinese Phys. C. \textbf{39} 093001 (2015).}
  
\bibitem{lumscan}
  M. Ablikim \textit{et al.} (BESIII Collaboration), \href{https://doi.org/10.1088/1674-1137/41/6/063001}{Chinese Phys. C. \textbf{41} 063001 (2017).}
    
\bibitem{detector}
  M. Ablikim \textit{et al.} (BESIII Collaboration), \href{https://doi.org/10.1016/j.nima.2009.12.050}{Nucl. Instrum. Methods Phys. Res., Sect. A \textbf{614}, 345, (2010).}
  
\bibitem{tof}
  X. Li \textit{et al.}, \href{https://doi.org/10.18429/JACoW-IPAC2016-TUYA01}{Proceedings of IPAC2016, Busan, Korea, 2016.}
  
\bibitem{geant4}
  S. Agostinelli \textit{et al.} (GEANT4 Collaboration), \href{https://doi.org/10.1016/S0168-9002(03)01368-8}{Nucl. Instrum. Methods Phys. Res., Sect. A \textbf{506}, 250 (2003).}


\bibitem{kkmc}
  S. Jadach, B. F. L. Ward, and Z. W\k{a}s, \href{https://doi.org/10.1103/PhysRevD.63.113009}{Phys. Rev. D \textbf{63}, 113009 (2001).}

\bibitem{evtgen}
  D. J. Lange, \href{https://doi.org/10.1016/S0168-9002(01)00089-4}{Nucl. Instrum. Methods Phys. Res., Sect. A \textbf{462}, 152 (2001).}

\bibitem{lund}                                                                       
  J.~C.~Chen, G.~S.~Huang, X.~R.~Qi, D.~H.~Zhang and Y.~S.~Zhu,                 
  \href{https://doi.org/10.1103/PhysRevD.62.034003}{Phys.\ Rev.\ D {\bf 62}, 034003 (2000)};
  R.~L.~Yang, R.~G.~Ping and H.~Chen,
  \href{https://doi.org/10.1088/0256-307X/31/6/061301}{Chin.\ Phys.\ Lett.\  {\bf 31}, 061301 (2014).}
  
\bibitem{photos}
  P. Golonka and Z. W\k{a}s, \href{https://doi.org/10.1140/epjc/s2005-02396-4}{Eur. Phys. J. C \textbf{45}, 97 (2006).}


\bibitem{Eichten:1974af}
E.~Eichten, K.~Gottfried, T.~Kinoshita, J.~B.~Kogut, K.~D.~Lane and T.~M.~Yan, \href{https://doi.org/10.1103/PhysRevLett.34.369}{Phys. Rev. Lett. \textbf{34}, 369-372 (1975) [erratum: Phys. Rev. Lett. \textbf{36}, 1276 (1976)]}
    
\bibitem{omegaChi}
  M. Ablikim \textit{et al.} (BESIII Collaboration), \href{https://doi.org/10.1103/PhysRevD.99.091103}{Phys. Rev. D. \textbf{99}, 091103(R) (2019).}

\bibitem{pipiPsi2S}
  M.Ablikim \textit{et al.} (BESIII Collaboration), \href{https://doi.org/10.1103/PhysRevD.104.052012}{Phys. Rev. D. \textbf{104}, 052012 (2021).}

  
\bibitem{yuping}
  M. Ablikim \textit{et al.} (BESIII Collaboration), \href{https://doi.org/10.1103/PhysRevD.87.012002}{Phys. Rev. D. \textbf{87}, 012002 (2013).}

\bibitem{Lara}
  M. Ablikim \textit{et al.} (BESIII Collaboration), \href{https://doi.org/10.1103/PhysRevD.96.012001}{Phys. Rev. D. \textbf{96}, 051101(R) (2017).}

\bibitem{lineshape}
  M. Ablikim \textit{et al.} (BESIII Collaboration), \href{https://doi.org/10.1103/PhysRevLett.118.092001}{Phys. Rev. Lett. \textbf{118}, 092001 (2017).}



\end{thebibliography}
\end{document}